\documentclass[a4paper,11pt]{article}
\pdfoutput=1
\usepackage{jheppub}
\usepackage{textcase}
\usepackage{graphicx,epsfig,psfrag,amssymb,hyperref}
\usepackage{multirow}
\usepackage{color,graphicx,epsfig,psfrag,amsmath,empheq}
\usepackage[dvipsnames]{xcolor}
\usepackage{bm}
\usepackage{mathrsfs,amsfonts,soul,color}
\usepackage[caption=false]{subfig}
\usepackage{hepunits}
\usepackage{enumitem}
\usepackage{placeins}
\usepackage{textcomp}

\usepackage{float}
\usepackage[toc,page]{appendix}
\usepackage[ruled,vlined]{algorithm2e}
\usepackage{wrapfig}
\SetKwInOut{Input}{Input}
\SetKwInOut{Output}{Output}
\SetKwFunction{Fforward}{Forward}
\SetKwFunction{Fbackward}{Backward}
\SetKwProg{Fn}{function}{:}{}
\DontPrintSemicolon

\begin{document}

\title{Simulation-based Anomaly Detection \\ for Multileptons at the LHC} 

\author{Katarzyna Krzy\.za\'nska$^{a}$}
\author{and Benjamin Nachman,$^{b,c}$}

\affiliation{
\phantom{ }\hspace{-0.12in}$^a$Department of Physics, Princeton University, Princeton, NJ 08544, USA \\
\phantom{ }\hspace{-0.12in}$^b$Physics Division, Lawrence Berkeley National Laboratory, Berkeley, CA 94720, USA \\
\phantom{ }\hspace{-0.12in}$^c$Berkeley Institute for Data Science, University of California, Berkeley, CA 94720, USA
}

\emailAdd{kk19@princeton.edu}
\emailAdd{bpnachman@lbl.gov}

\abstract{
Decays of Higgs boson-like particles into multileptons is a well-motivated process for investigating physics beyond the Standard Model (SM).  A unique feature of this final state is the precision with which the SM is known.  As a result, simulations are used directly to estimate the background.  Current searches consider specific models and typically focus on those with a single free parameter to simplify the analysis and interpretation. In this paper, we explore recent proposals for signal model agnostic searches using machine learning in the multilepton final state.  These tools can be used to simultaneously search for many models, some of which have no dedicated search at the Large Hadron Collider.  We find that the machine learning methods offer broad coverage across parameter space beyond where current searches are sensitive, with a necessary loss of performance compared to dedicated searches by only about one order of magnitude.
}

\maketitle

\section{Introduction}
\label{sec:intro}

Many models of new physics contain multiple free parameters that are not predicated by the theory.  Typically, searches pick a one- or two-dimensional slice through this space to set limits on model parameters (see Ref.~\cite{atlasexoticstwiki,atlassusytwiki,atlashdbspublictwiki,cmsexoticstwiki,cmssusytwiki,cmsb2gtwiki,lhcbtwiki}).  This is an effective way to visualize results, but it can also hide potential anomalous regions of the parameter space that are not well-captured by particular slices.  In the usual approach, a small number of \textit{signal regions} (SRs) are constructed to achieve broad sensitivity over the low-dimensional slices through parameter space.  These SRs are often optimized using benchmark models.  In order to improve sensitivity, SR definitions are sometimes parameterized by a subset of the BSM parameters.  Scanning through the parameter space in this way will improve the sensitivity to particular models away from the benchmark models at the cost of analysis complexity and a larger trials factor.

An alternative search strategy is \textit{anomaly detection} (AD).  AD searches reduce signal or background model dependence by directly training with data.  As real data do not have labeled examples (signal versus background) for analysis optimization, these analysis analysis techniques are called \textit{less than supervised}.  This paper will focus on signal model-agnostic AD methods.  A variety of classical AD searches of this type have been conducted at D0~\cite{sleuth,Abbott:2000fb,Abbott:2000gx,Abbott:2001ke}, H1~\cite{Aaron:2008aa,Aktas:2004pz}, ALEPH~\cite{Cranmer:2005zn}, CDF~\cite{Aaltonen:2007dg,Aaltonen:2007ab,Aaltonen:2008vt}, CMS~\cite{CMS-PAS-EXO-14-016,CMS-PAS-EXO-10-021,CMS:2020ohc,Sirunyan:2020jwk}, and ATLAS~\cite{Aaboud:2018ufy,ATLAS-CONF-2014-006,ATLAS-CONF-2012-107}.

Recently, there have been many proposals for automating AD methods with machine learning~\cite{DAgnolo:2018cun,Collins:2018epr,Collins:2019jip,DAgnolo:2019vbw,Andreassen:2020nkr,Nachman:2020lpy,Hallin:2021wme,Farina:2018fyg,Heimel:2018mkt,Roy:2019jae,Cerri:2018anq,Blance:2019ibf,Hajer:2018kqm,DeSimone:2018efk,Mullin:2019mmh,1809.02977,Dillon:2019cqt,Aguilar-Saavedra:2017rzt,Romao:2019dvs,Romao:2020ojy,knapp2020adversarially,collaboration2020dijet,1797846,1800445,Amram:2020ykb,Cheng:2020dal,Khosa:2020qrz,Thaprasop:2020mzp,Alexander:2020mbx,aguilarsaavedra2020mass,1815227,pol2020anomaly,Mikuni:2020qds,vanBeekveld:2020txa,Park:2020pak,Faroughy:2020gas,Stein:2020rou,Chakravarti:2021svb,Batson:2021agz,Blance:2021gcs,Bortolato:2021zic,Collins:2021nxn,Dillon:2021nxw,Finke:2021sdf,Shih:2021kbt,Atkinson:2021nlt,Kahn:2021drv,Dorigo:2021iyy,Caron:2021wmq,Govorkova:2021hqu,Kasieczka:2021tew,Volkovich:2021txe,Govorkova:2021utb,Ostdiek:2021bem,Fraser:2021lxm,Chakravarti:2021svb,Kasieczka:2021xcg,Aarrestad:2021oeb} (see Ref.~\cite{Kasieczka:2021xcg,Aarrestad:2021oeb,2112.03769,Feickert:2021ajf} for overviews of the field).  One particularly sensitive class of signal model agnostic searches train classifiers to directly distinguish data in a particular region of phase space (SR) from a prediction of the SM background~\cite{Collins:2018epr,Collins:2019jip,DAgnolo:2018cun,DAgnolo:2019vbw,Andreassen:2020nkr,Nachman:2020lpy,Hallin:2021wme,Chakravarti:2021svb,dAgnolo:2021aun,Hallin:2021wme}.  These searches are called weakly (or semi) supervised because one dataset has known labels (background) while one dataset has noisy labels (background + maybe signal).

These weakly supervised methods are distinguished by the dataset used to estimate the background.  If the background is well-understood theoretically, then the reference sample could be simulation~\cite{DAgnolo:2018cun,DAgnolo:2019vbw,dAgnolo:2021aun,Chakravarti:2021svb}. This has the advantage that the background prediction does not need to be learned, but has the disadvantage of being strongly background-model dependent.  There are few final states at the LHC for which the background is known precisely enough to be used directly for background estimation.  One exception is the final state with four charged leptons.  Both ATLAS~\cite{ATLAS:2020tlo,ATLAS:2018coo,ATLAS:2020wny,ATLAS:2021ldb} and CMS~\cite{CMS:2016ilx,CMS:2020bni,cmscollaboration2021search,CMS:2021nnc} directly use Monte Carlo (MC) simulations to estimate the background and ATLAS even uses machine learning to isolate particular signals~\cite{ATLAS:2020tlo}.  While powerful, this approach is signal model-specific and does not readily extend to models with multidimensional parameters.  In this paper, we explore AD methods applied to the four lepton final state.  In particular, we train classifiers to distinguish (simulated) data from background predictions.  This approach is complementary to direct searches and has broad sensitivity in a signal model-agnostic approach. 

This paper is organized as follows.  Section~\ref{sec:sim} introduces the simulated samples used for the machine learning studies.  The machine learning methods are introduced in Sec.~\ref{sec:methods} and numerical results are presented in Sec.~\ref{sec:results}. The paper ends with conclusions and outlook in Sec.~\ref{sec:concl}.

\section{Simulations}
\label{sec:sim}

The generation of background and signal events is performed with \textsc{MadGraph5\_aMC@NLO} 2.8.0~\cite{Alwall:2014hca}.  The signal is generated using the Higgs Effective Field Theory (\texttt{heft}) via \verb#p p > h# with a variable Higgs mass and the background is generated also using \texttt{heft} via \verb#p p > e+ e- mu+ mu-#.  In principle, the same analysis could be applied in the 4-electron and 4-muon final states, but additional complication arises from combinatorical factors.  We leave the exploration of such final states to future work.  Simulated events are passed to decay (signal only), parton showering, and hadronization using \textsc{Pythia}~8.244~\cite{Sjostrand:2006za,Sjostrand:2007gs,Sjostrand:2014zea} with its default settings. The detector simulation is parameterized with \textsc{Delphes} 3.4.2~\cite{deFavereau:2013fsa,Mertens:2015kba,Selvaggi:2014mya} using the default CMS card.  In what follows, we will assume that the background is known without any systematic uncertainties and so the `data' will be an independent, but statistically identical copy of the SM background.  See Ref.~\cite{dAgnolo:2021aun} for a discussion of how this could be extended to include systematic uncertainties.  The number of background events is chosen to match\footnote{In principle, the simulation could have a larger number of events than in `data'.  Studies in Ref.~\cite{Hallin:2021wme} suggest this could be effective in the case of surrogate models (not simulation).  We leave a detailed study of the relative size of the simulation dataset in to future studies.} the LHC Run 2 dataset of about 150 fb$^{-1}$.

Focusing only on the leptons\footnote{Other event properties could also be useful for discrimination.  However, information about the hadronic final state is known with less precision and thus may introduce the need to involve data-driven background estimation.}, each event is characterized by 12 numbers (four three-momenta).  For many models of the form $pp\rightarrow A\rightarrow B(\rightarrow e^+e^-)C(\rightarrow \mu^+\mu^-)$, the three masses $m_{e^+e^-\mu^+\mu^-},m_{e^+e^-},m_{\mu^+\mu^-}$ are nearly sufficient statistics for characterizing the new physics.  In this paper, we consider signals of this form, where $A$ is a variable-mass Higgs boson that decays to two different mass scalars.  For this reason, we focus on the three-dimensional problem in this paper.  Non-resonant signals and signals with non-trivial spin structures could benefit from using more of the phase space.

\begin{figure}[h!]
    \centering
    \includegraphics[width=0.45\textwidth]{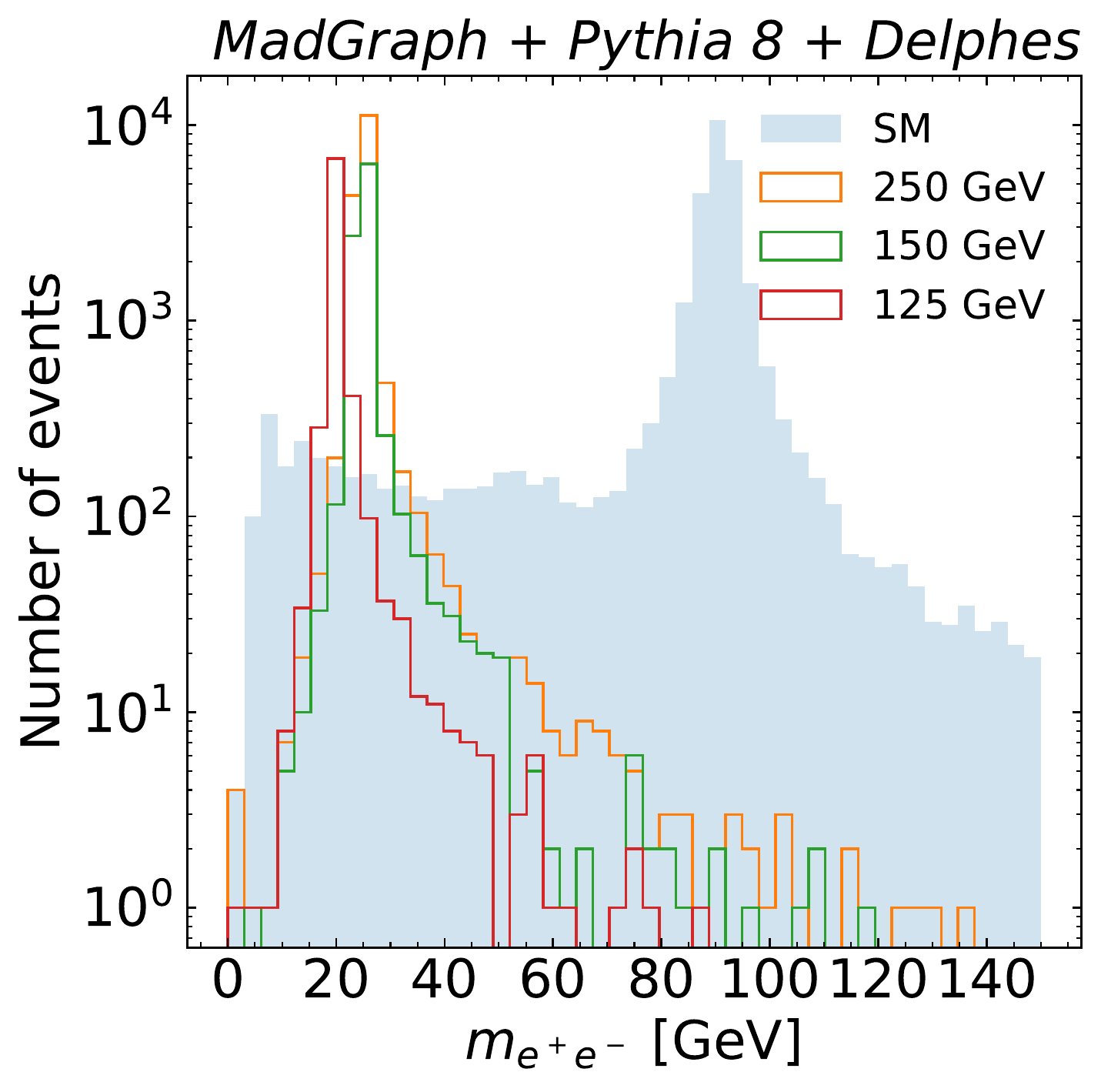}\includegraphics[width=0.45\textwidth]{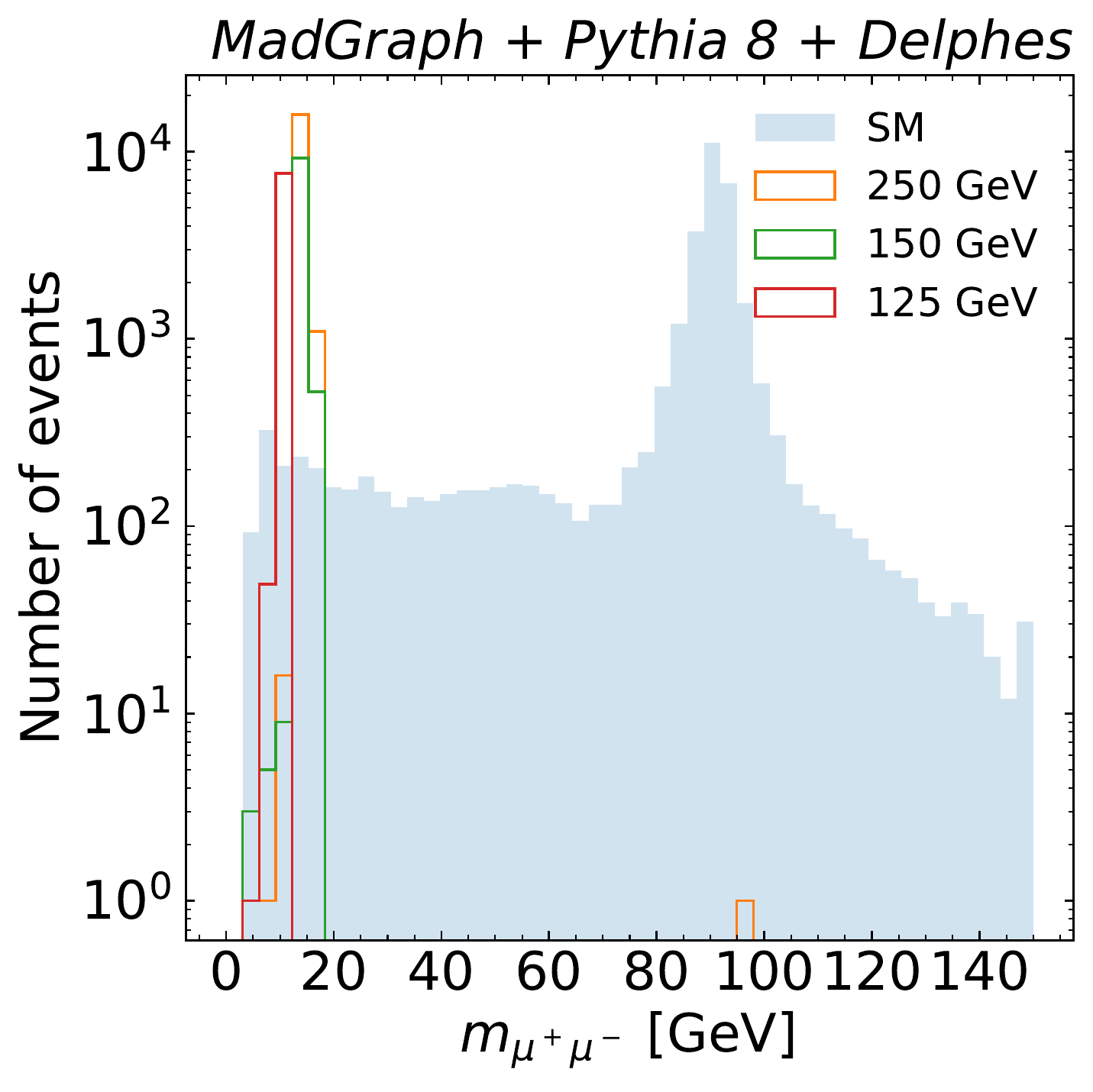}\\\includegraphics[width=0.45\textwidth]{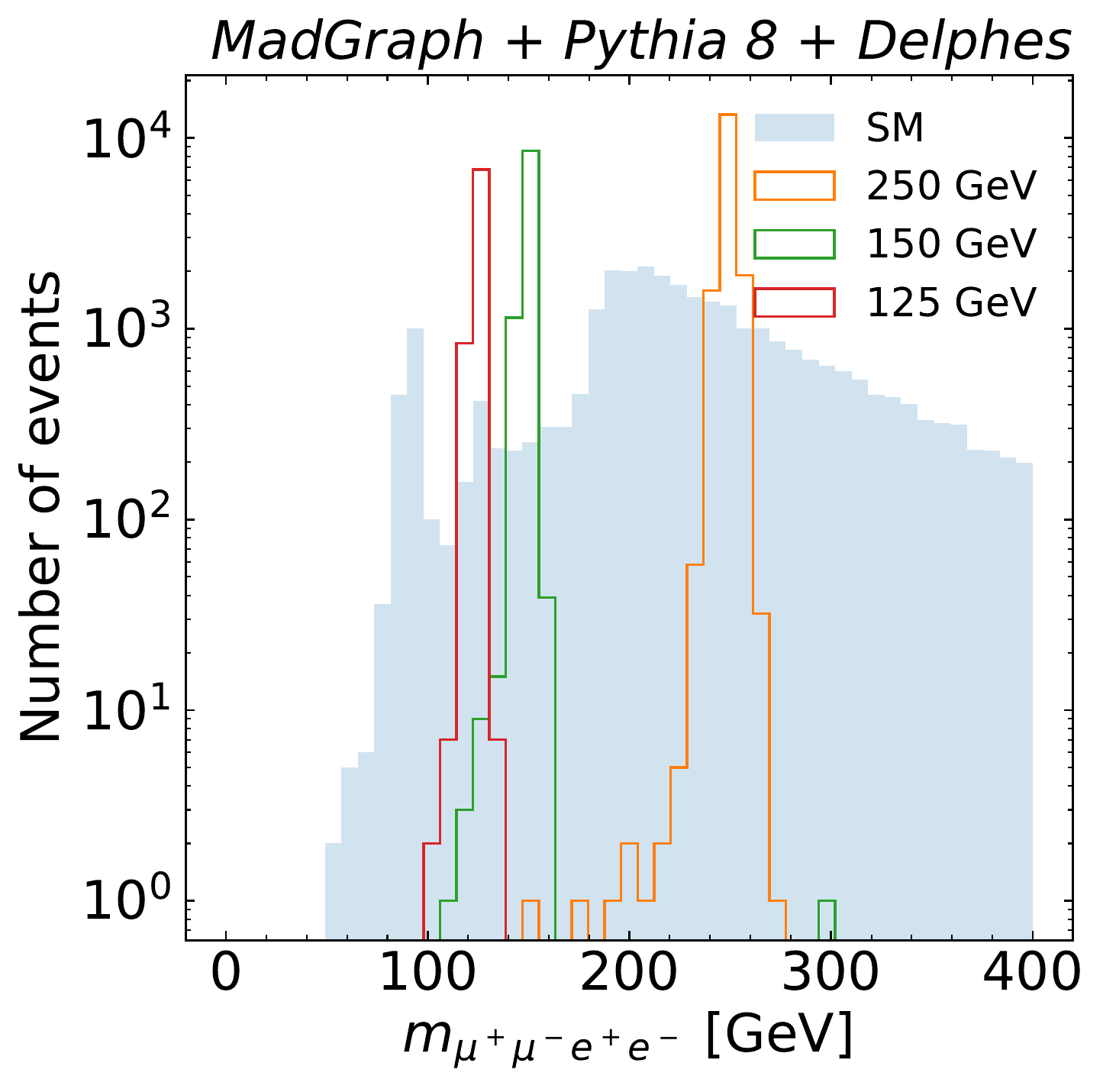}
    \caption{The three dimensions used for machine learning: $m_{e^+e^-}$ (left), $m_{\mu^+\mu^-}$ (right) and $m_{4\ell}$ (bottom). }
    \label{fig:inputs}
\end{figure}

The spectra of the three invariant masses for the background and three representative signals are presented in Fig.~\ref{fig:inputs}.  As expected, the di-electron and di-muon invariant masses peak near the $Z$ boson mass of 90 GeV~\cite{Zyla:2020zbs} and there are peaks in the four-lepton invariant mass at the $Z$ peak and the Higgs boson mass of about 125 GeV~\cite{Zyla:2020zbs}.  Each of the signals is resonant in all three observables with peaks at the masses of the particles.  In particular, our three signal models have parent masses of 125, 150, and 250 GeV, respectively.  In each case, the parent particle decays to two children, with masses of 25 and 15 GeV for electrons and muons, respectively.  These parameters are chosen to be representative; the resulting less-than-supervised analysis does not rely on them in the machine learning training.

\section{Methods}
\label{sec:methods}

Our machine learning approach is simple, but powerful: we train a classifier to distinguish SM background from `data'.  This `data' is SM background with some amount of injected signal.  If the classifier is able to significantly differentiate these samples, then there is evidence that the `data' contains BSM events.  The significance is determined via bootstrapping~\cite{efron1979}.  In particular, we create $N$ bootstrap datasets by sampling (with replacement) from the SM background.  We then train $N$ classifiers, where each one distinguishes the bootstrap `data' from the SM background.  We compare the classifier performance in the nominal case with the distribution from the bootstraps to compute $p$-values.  

There is no unique choice for which statistic to use to evaluate the performance.  Theoretically, the optimal statistics in the absence of systematic uncertainties are those that are monotonically related to the likelihood ratio\footnote{See also Appendix A in Ref.~\cite{Nachman:2020lpy}.}~\cite{neyman1933ix}.  In a sense, we are performing a multidimensional, unbinned, and nearly non-parameteric\footnote{We will use neural networks for the machine learning; these are by definition parameterized functions, but there are so many parameters that they are effectively non-parametric.} goodness of fit.  It may be possible to exploit asymptotic formulae to avoid bootstrapping with additional assumptions~\cite{DAgnolo:2018cun}, but we focus on the numerical approach because it is more general.  See Ref.~\cite{Chakravarti:2021svb} for a comparison of bootstrapping to other approaches.

The classifiers are parameterized as fully connected deep neural networks\footnote{Given the low-dimensionality of our demonstration, a shallower classifier such as a Boosted Decision Tree would likely also be effective.  However, neural network training is straight forward and naturally extends to higher dimensions.}.  These networks are implemented in \textsc{TensorFlow}~\cite{tensorflow} 2.2.0 and \textsc{Keras}~\cite{keras} and optimized using \textsc{Adam}~\cite{adam}.  The networks have three hidden layers with 50 nodes per layer.  Rectified linear activation functions are used for all intermediate layers.   We consider two loss functions ($L$):

\begin{itemize}
    \item Binary Cross Entropy Loss (BCE).  This is the most widely-used loss function in machine learning:
    
    \begin{align}
        L[f]=-\sum_\text{$x\in$ `data'} \log(f(x)) - \sum_\text{$x\in$ background} \log(1-f(x))\,,
    \end{align}
    where the last layer of the neural network is a sigmoid so that $0\leq f\leq 1$.
    
    \item Maximum Likelihood Classifier Loss~\cite{DAgnolo:2018cun,2101.07263} (MLC).  This loss function directly learns the (log) likelihood ratio, which results in useful asymptotic statistical properties:
    
    \begin{align}
        L[f]=-\sum_\text{$x\in$ `data'} (f(x)-1) - \sum_\text{$x\in$ background} e^{f(x)}\,,
    \end{align}
    where the last layer of the neural network is now linear so $-\infty < f < \infty$.    
    
\end{itemize}
All of the inputs to the neural networks are standardized to have zero mean and unit variance.  In order to improve performance, we train 10 neural networks with different random initialization and take the average value per event.  The networks are each trained for a fixed 20 epochs.  None of the network or training parameters were highly optimized.

In addition to comparing BCE and MLC, we also explore multiple metrics for performing the hypothesis test.  In particular, we consider the average loss, the average score, the maximum score, and the standard deviation over scores.  The average loss is the quantity that is optimized.  By averaging over the full phase space, it could be that some information about localized anomalies is lost.  This is the motivation for examining also the maximum score and the standard deviation across scores.  These quantities are more sensitive to localized deviations, but are also more sensitive to background fluctuations.  See also Ref.~\cite{Chakravarti:2021svb} for studies with alternative metrics, such as area under the receiver operating characteristic (ROC) curve.

\section{Results}
\label{sec:results}

\begin{figure}
    \centering
    \includegraphics[width=0.45\textwidth]{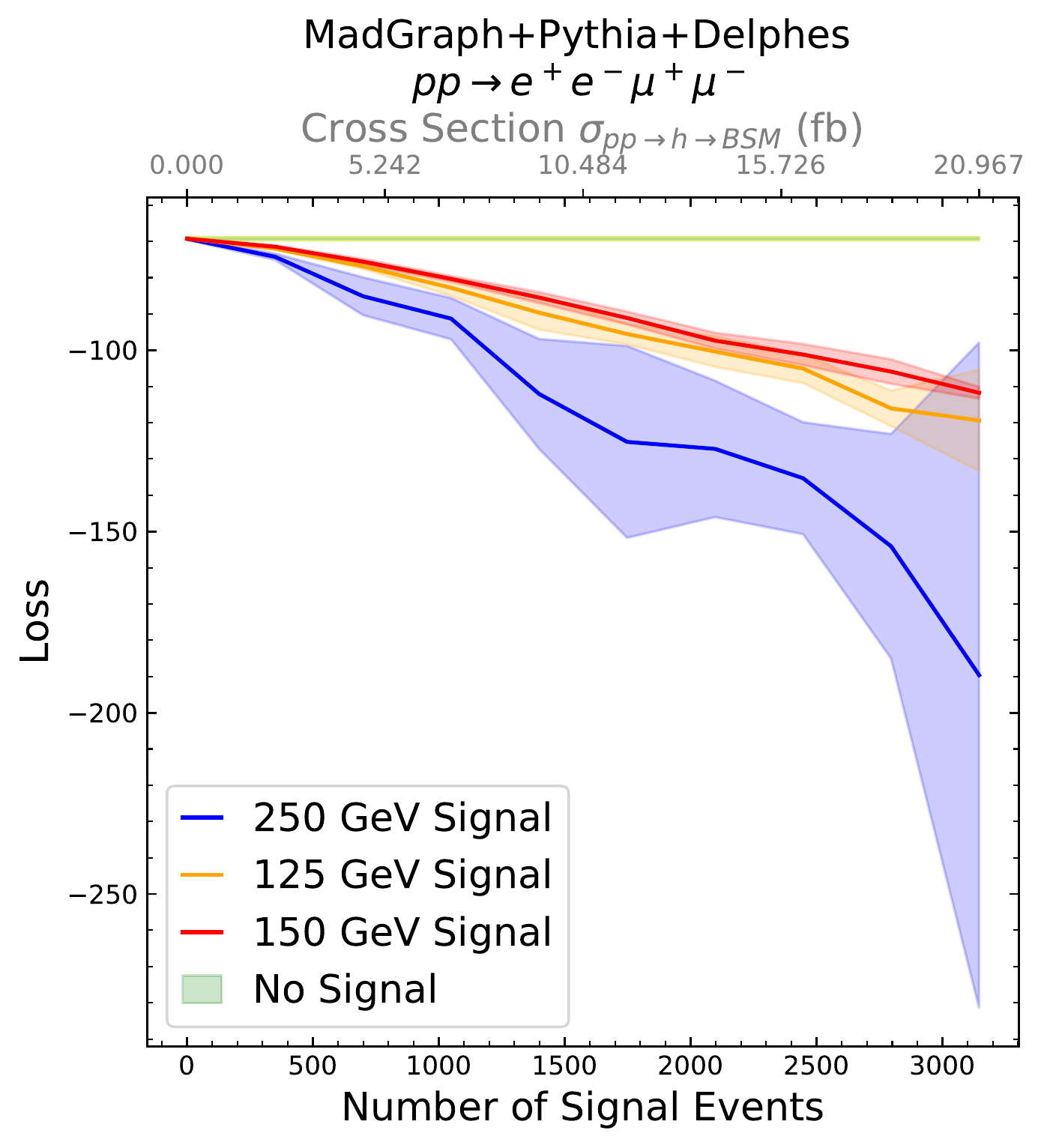}\includegraphics[width=0.45\textwidth]{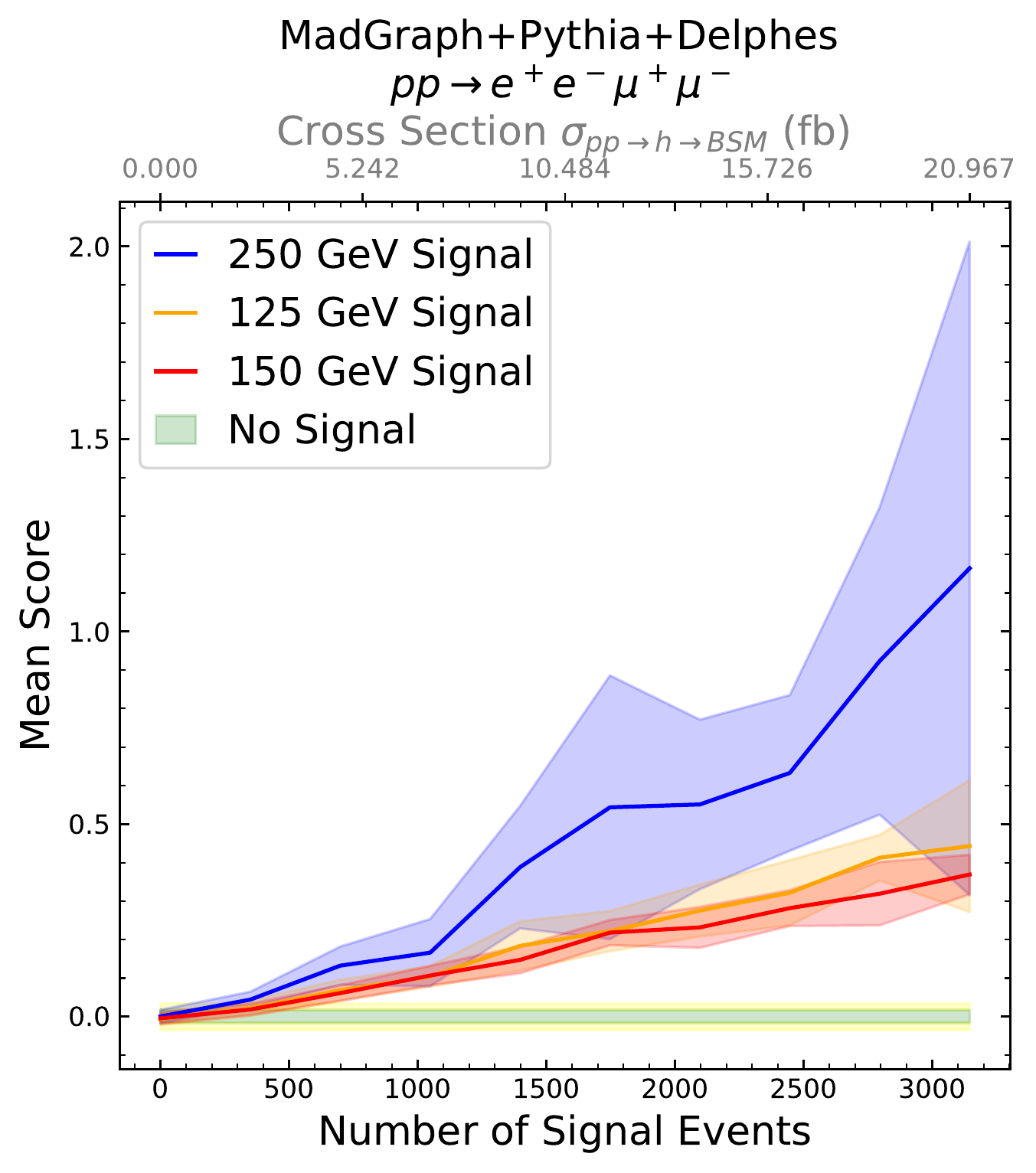}\\
    \includegraphics[width=0.45\textwidth]{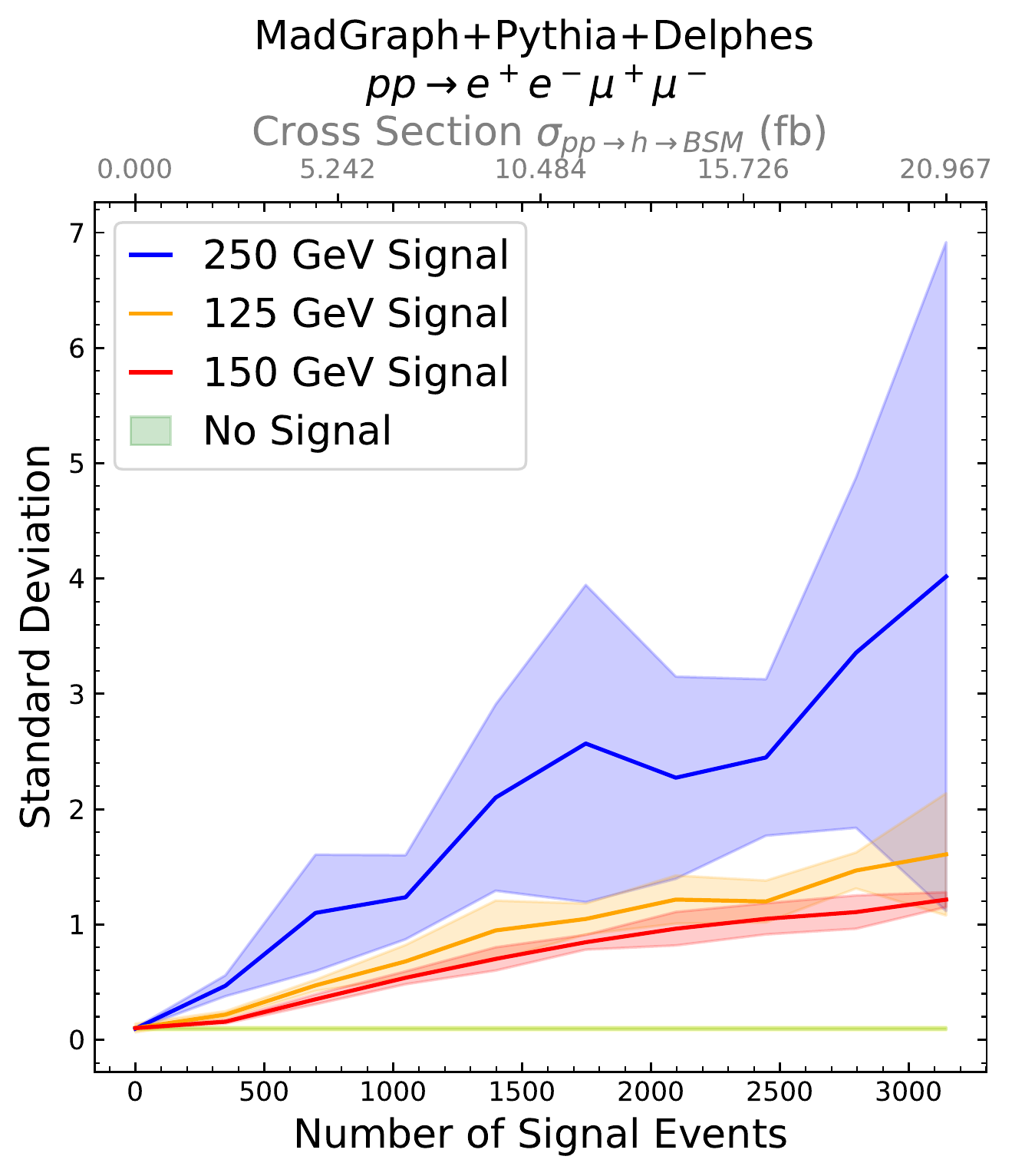}\includegraphics[width=0.45\textwidth]{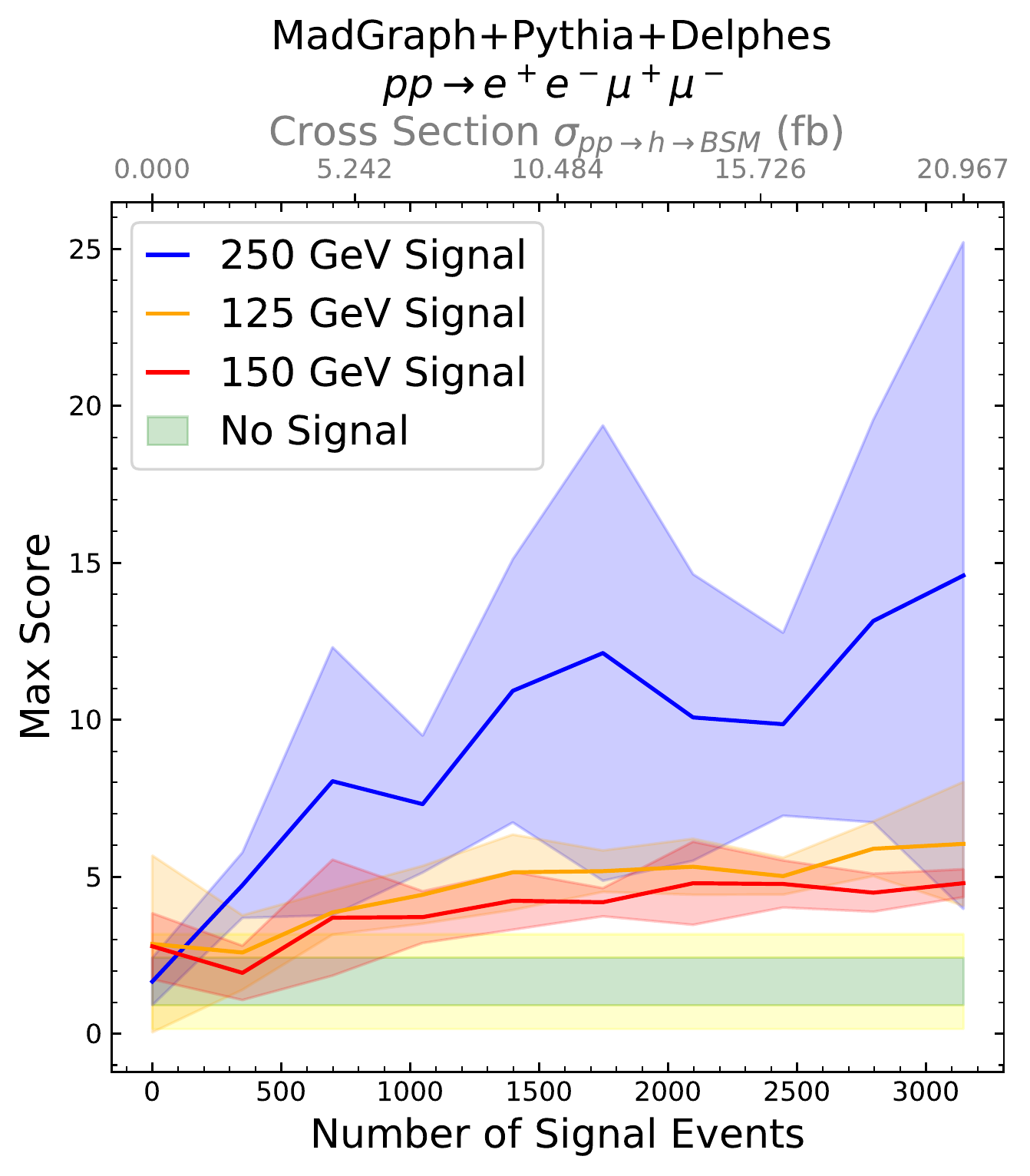}
    \caption{\textbf{MLC Loss}.  The average loss (top left), average score (top right), score standard deviation (lower left), and maximum score (lower right) when using the MLC loss as a function of the number of injected signals.  The green (yellow) band corresponds to the $1\sigma$ (2$\sigma$) region of the background-only hypothesis.  The bands for each signal correspond to the standard deviation across 10 independent signal injections.  The upper axis provides the injected cross section, where the number of background events corresponds to 150 fb$^{-1}$.}
    \label{fig:MLC_zoomout}
\end{figure}

\begin{figure}
    \centering
    \includegraphics[width=0.45\textwidth]{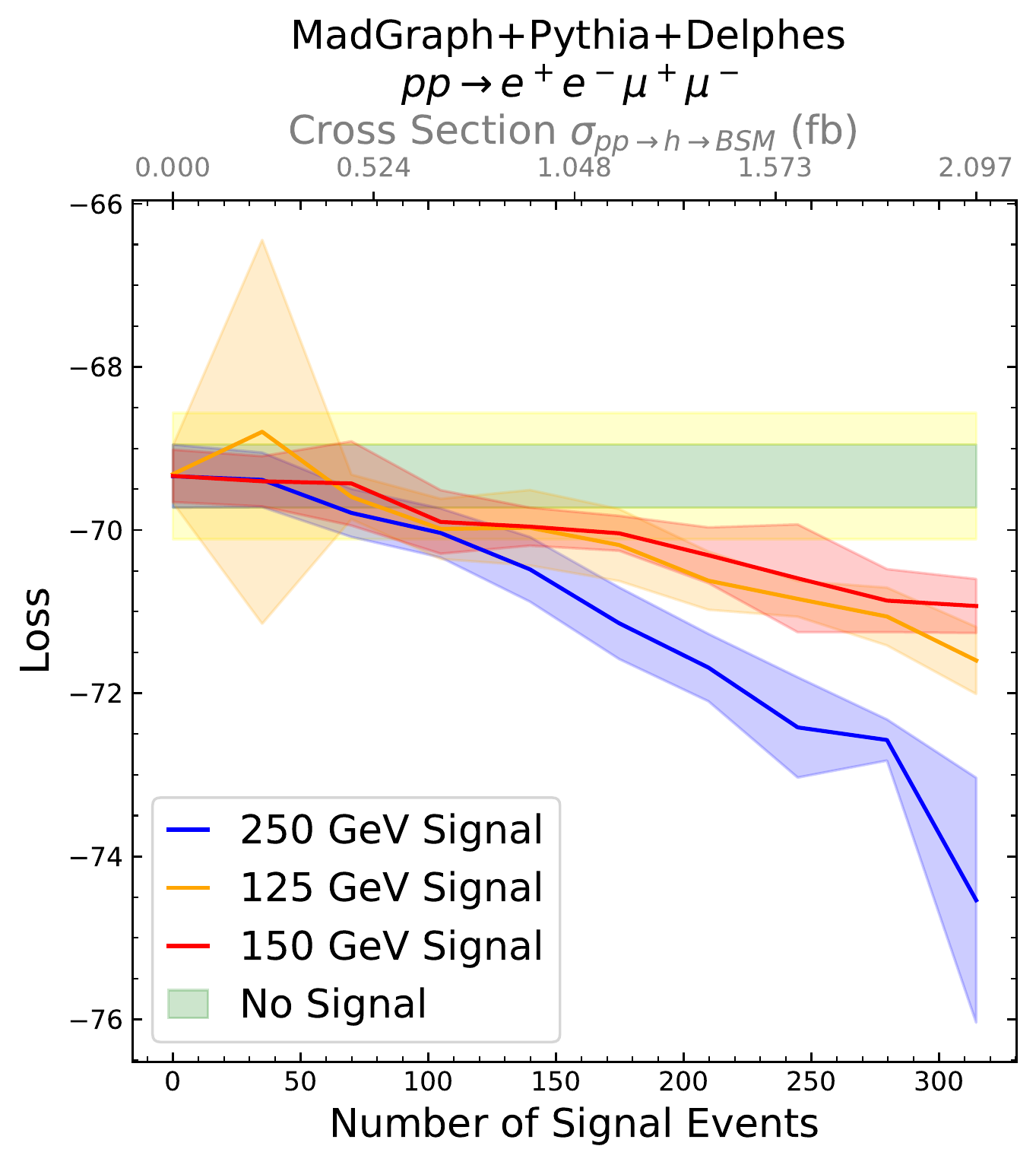}\includegraphics[width=0.45\textwidth]{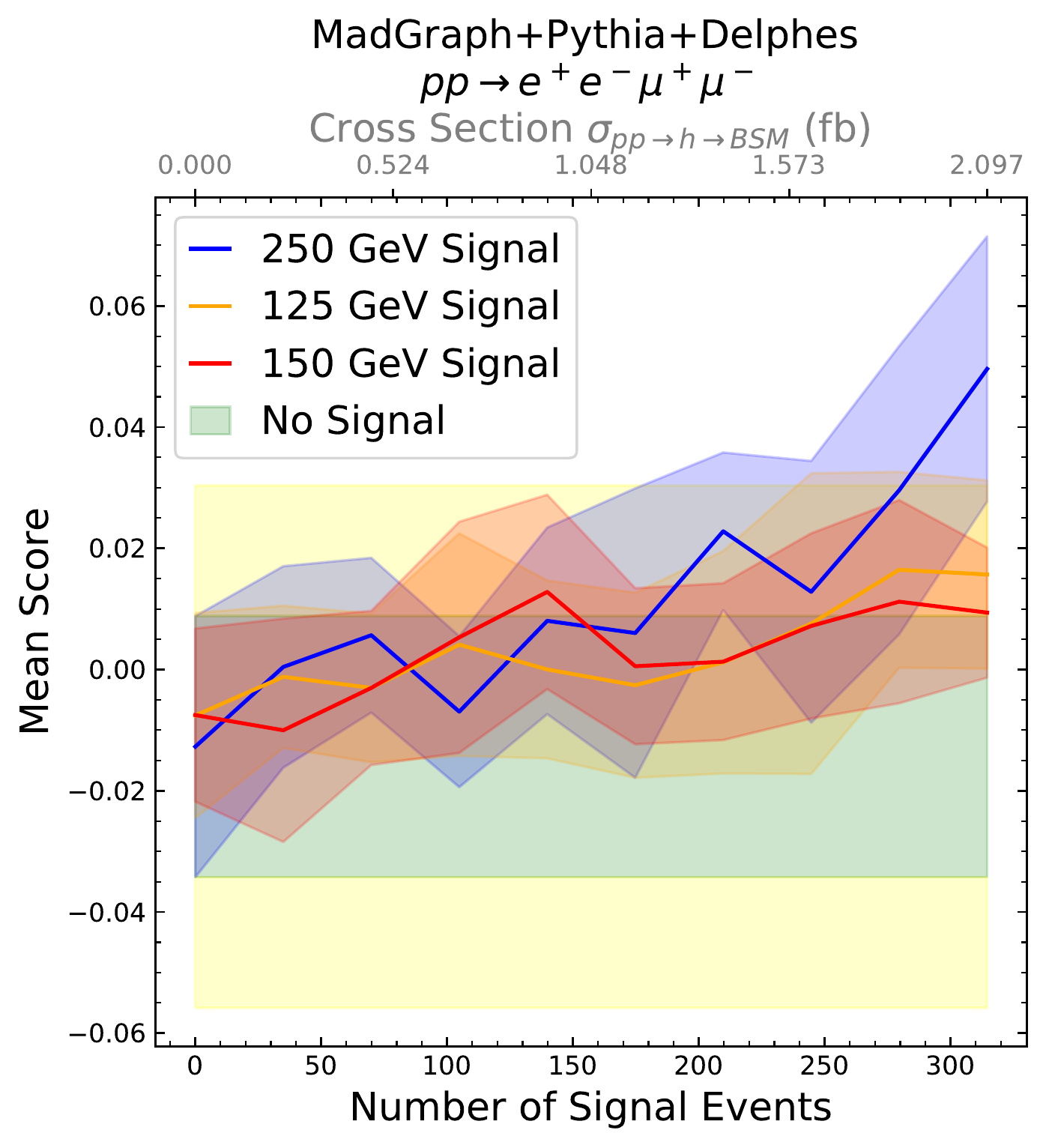}\\
    \includegraphics[width=0.45\textwidth]{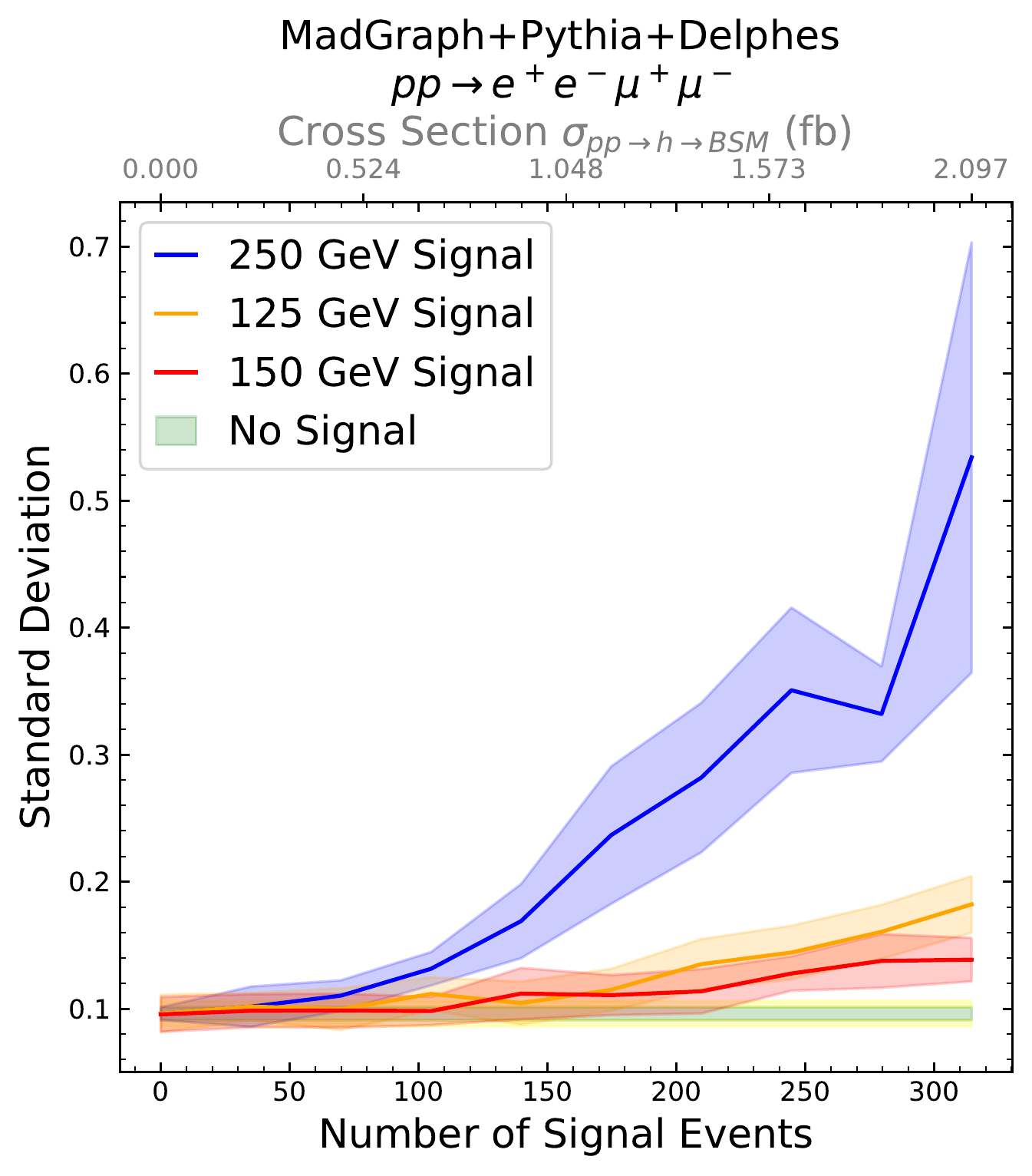}\includegraphics[width=0.45\textwidth]{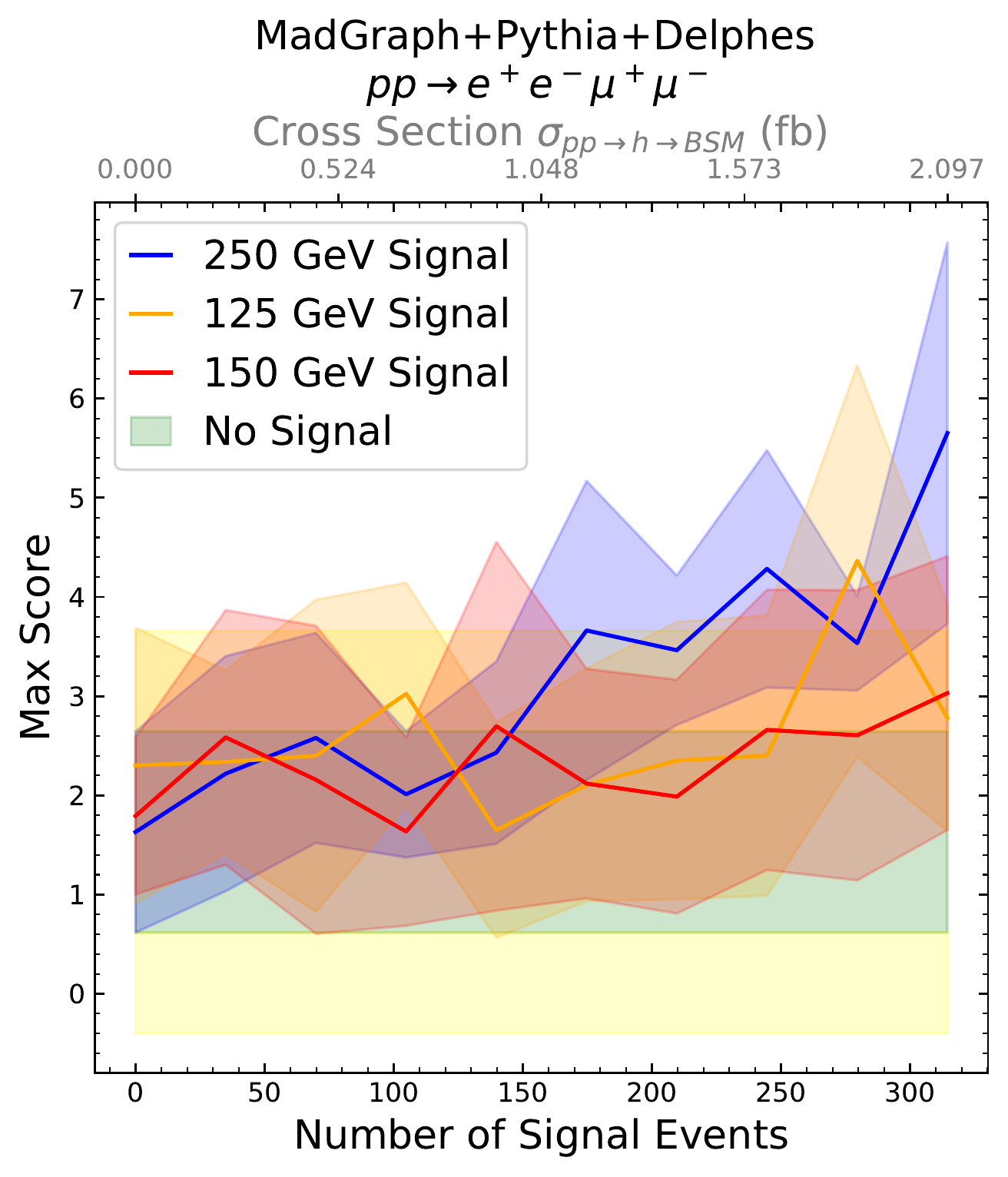}
    \caption{\textbf{MLC Loss}.  Same as Fig.~\ref{fig:MLC_zoomout}, but zoomed in by a factor of 10 on the horizontal axis.  The 95\% exclusion limits approximately correspond to where the lines for each signal cross the yellow $2\sigma$ band for the background-only hypothesis.}
    \label{fig:MLC_zoomin}
\end{figure}

\begin{figure}
    \centering
    \includegraphics[width=0.45\textwidth]{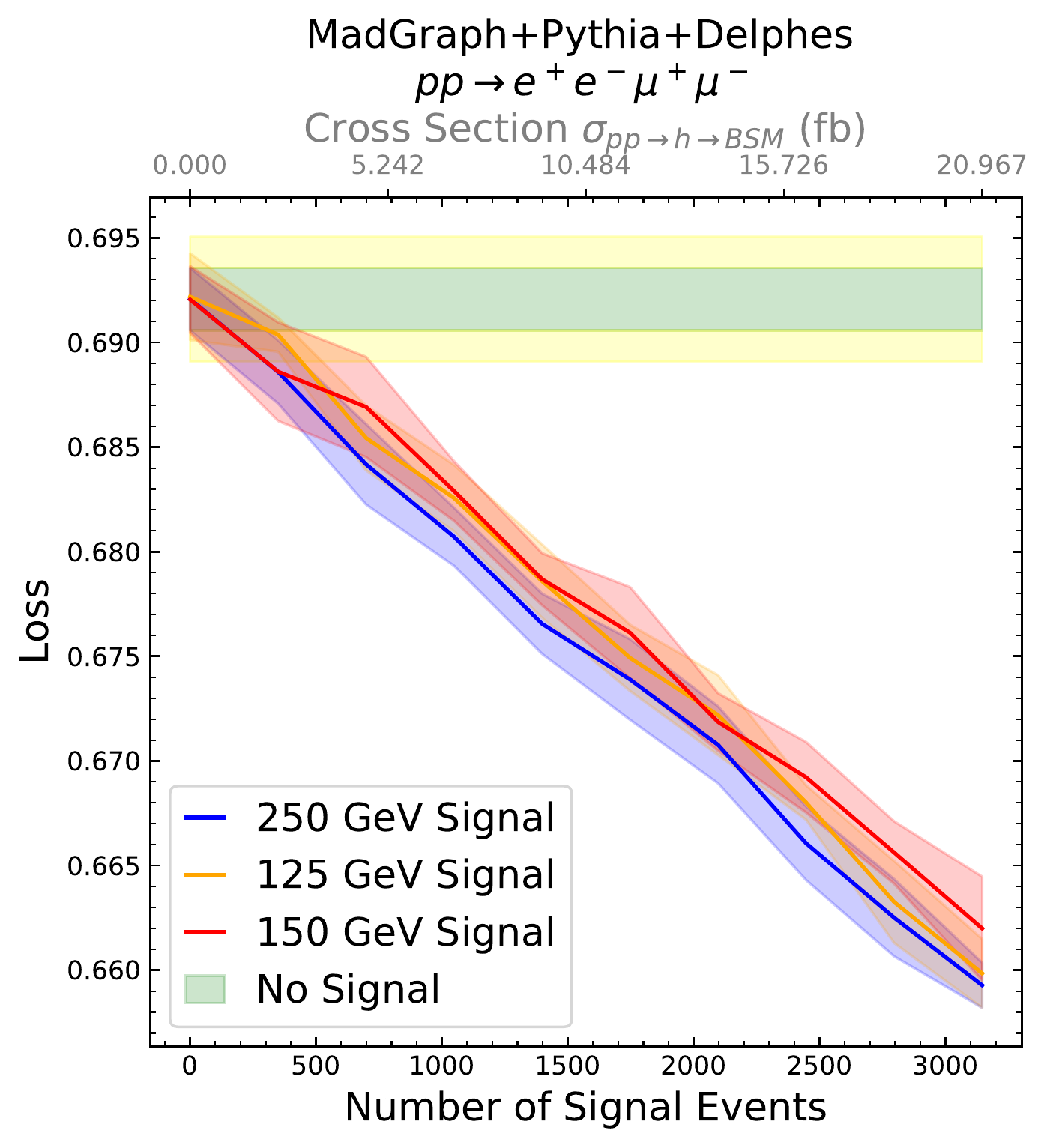}\includegraphics[width=0.45\textwidth]{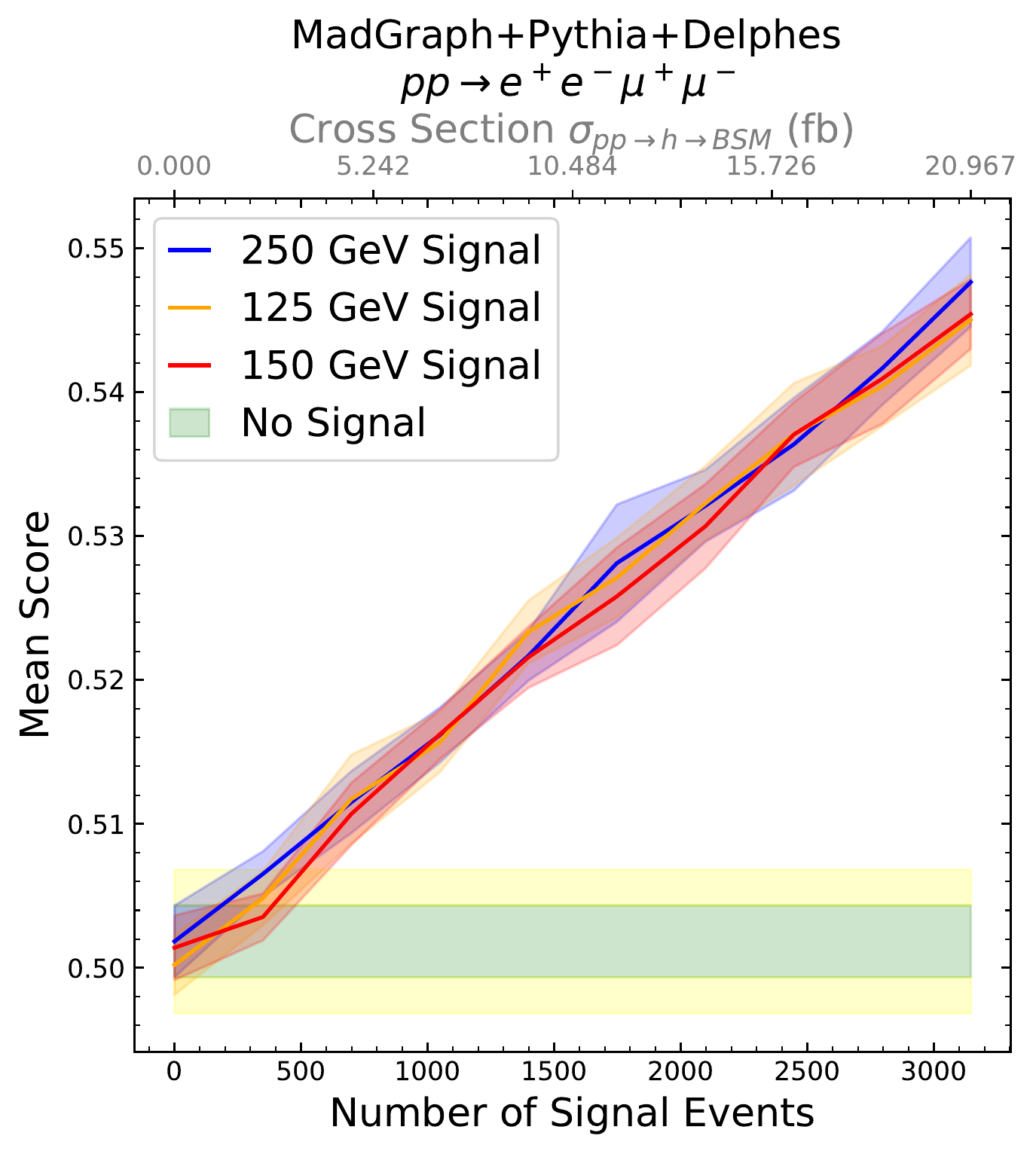}\\
    \includegraphics[width=0.45\textwidth]{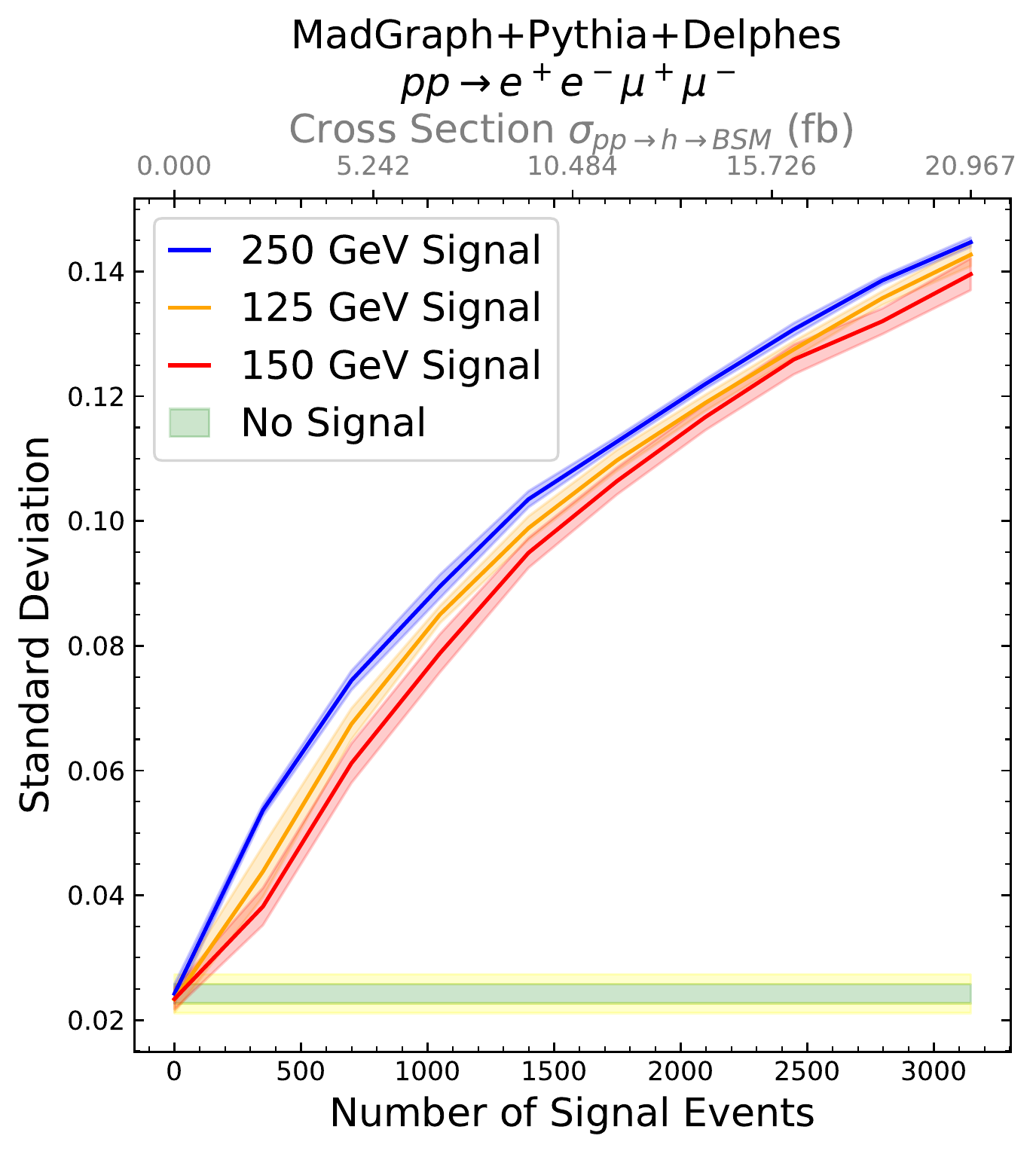}\includegraphics[width=0.45\textwidth]{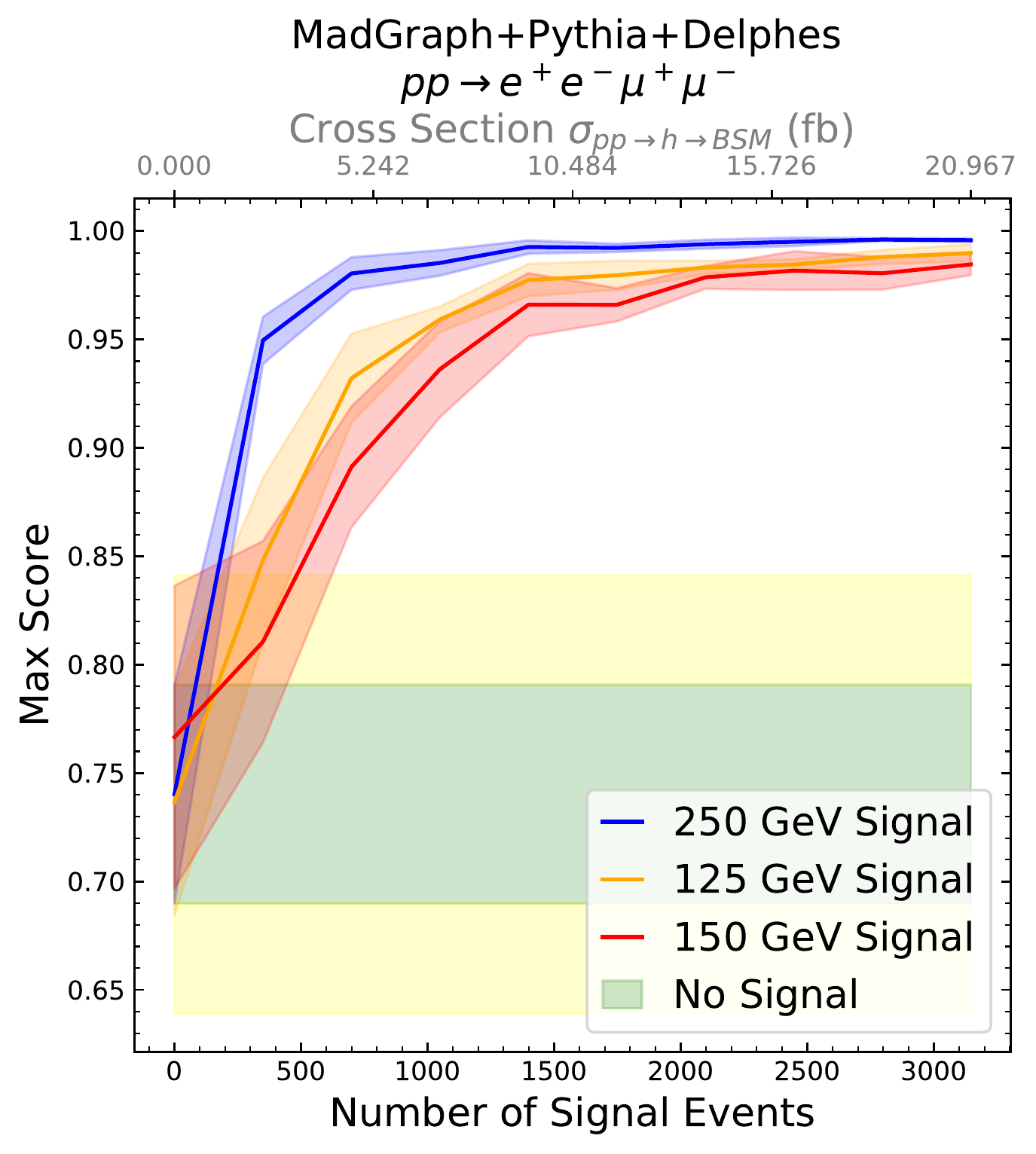}
    \caption{\textbf{BCE Loss}.  The average loss (top left), average score (top right), score standard deviation (lower left), and maximum score (lower right) when using the MLC loss as a function of the number of injected signals.  The green (yellow) band corresponds to the $1\sigma$ (2$\sigma$) region of the background-only hypothesis.  The bands for each signal correspond to the standard deviation across 10 independent signal injections.  The upper axis provides the injected cross section, where the number of background events corresponds to 150 fb$^{-1}$.}
    \label{fig:BCE_zoomout}
\end{figure}

\begin{figure}
    \centering
    \includegraphics[width=0.45\textwidth]{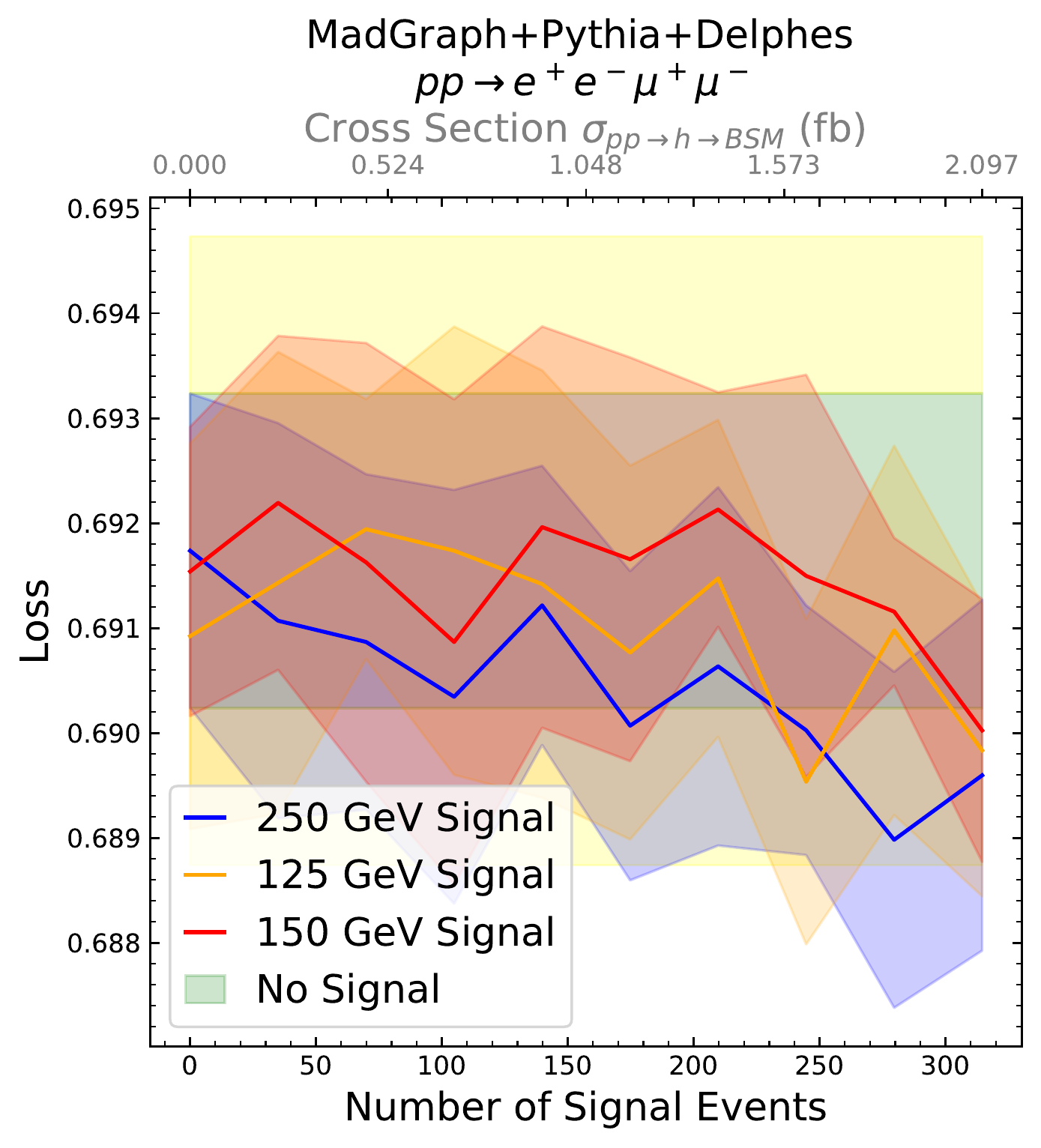}\includegraphics[width=0.45\textwidth]{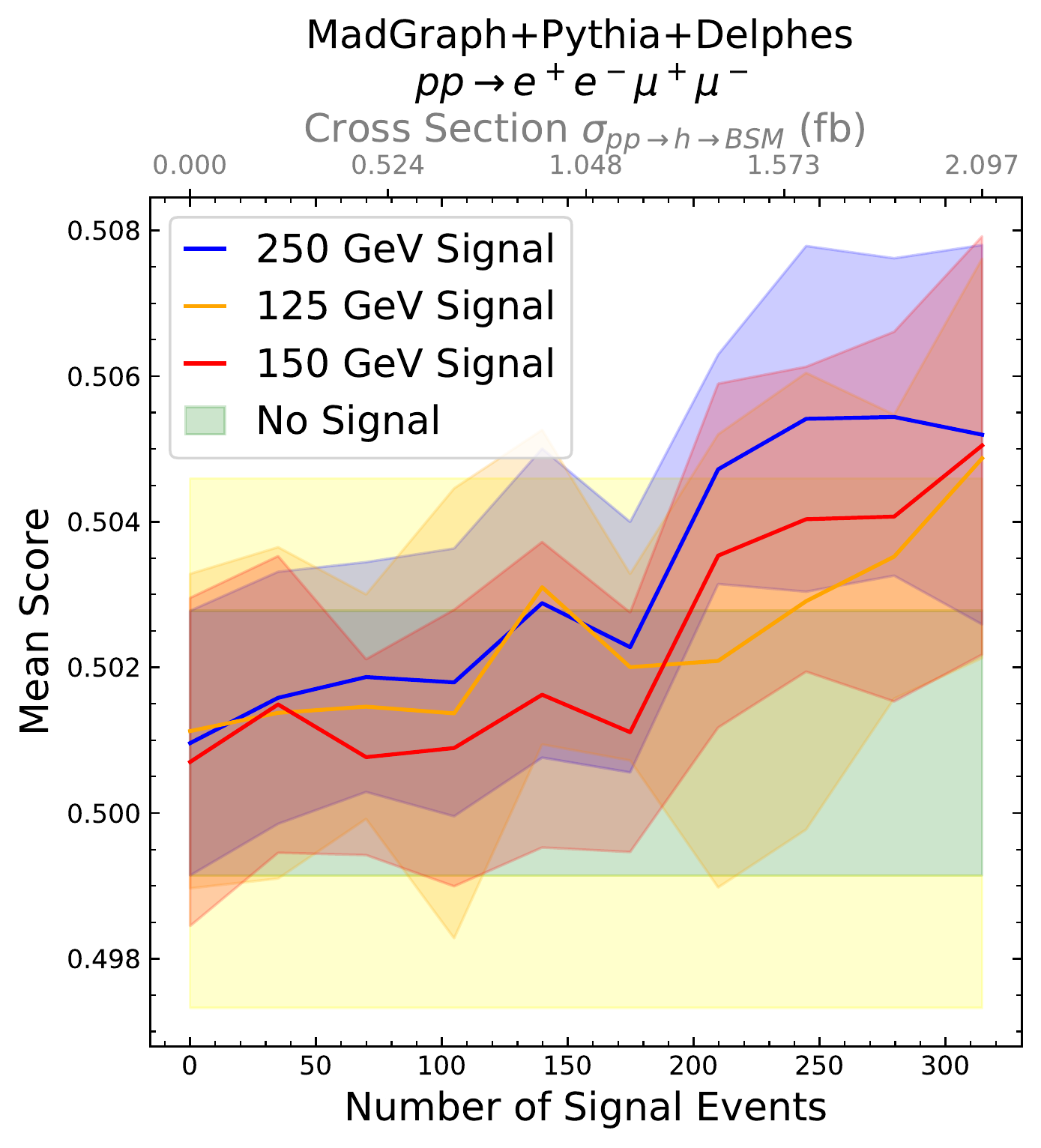}\\
    \includegraphics[width=0.45\textwidth]{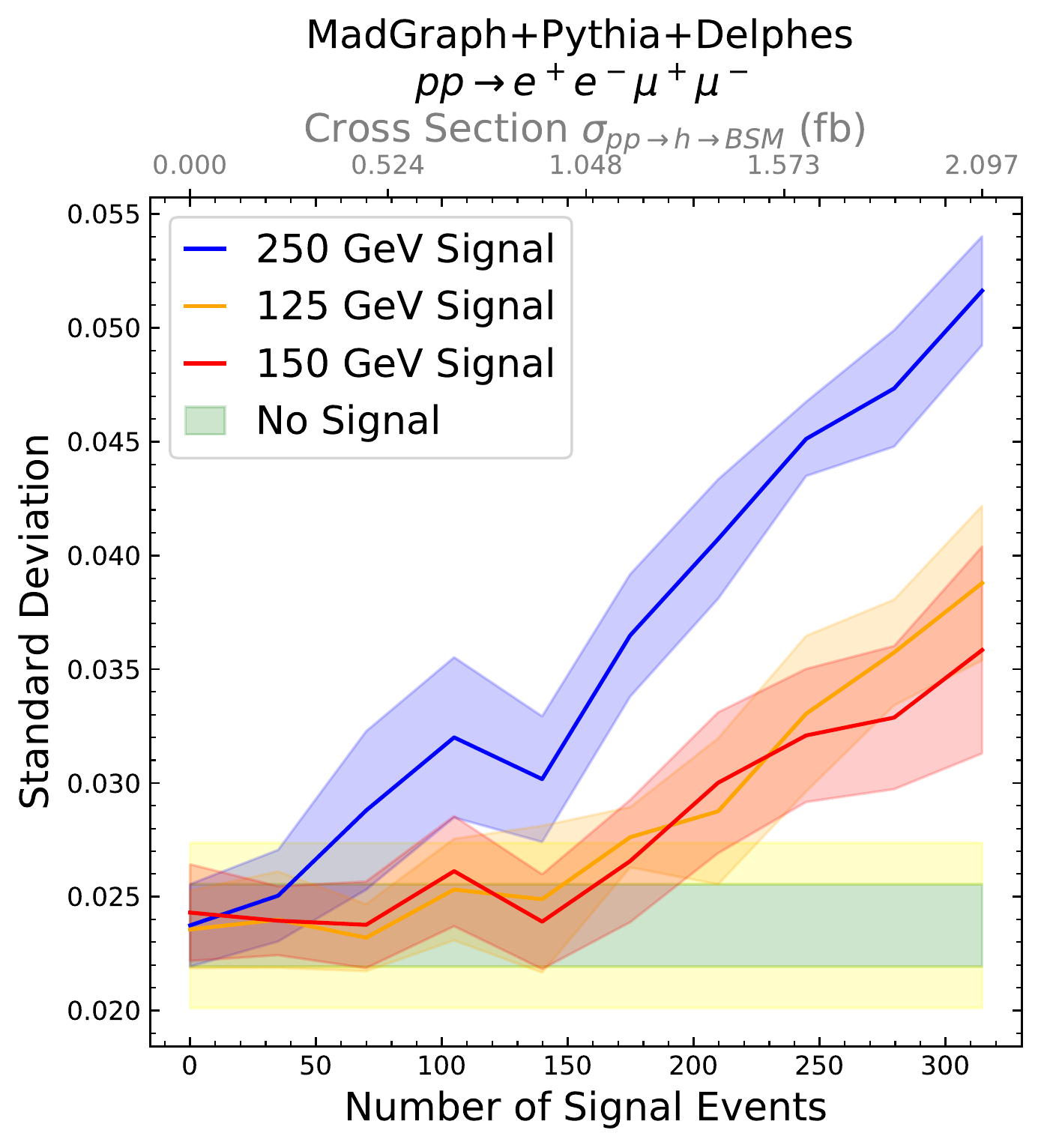}\includegraphics[width=0.45\textwidth]{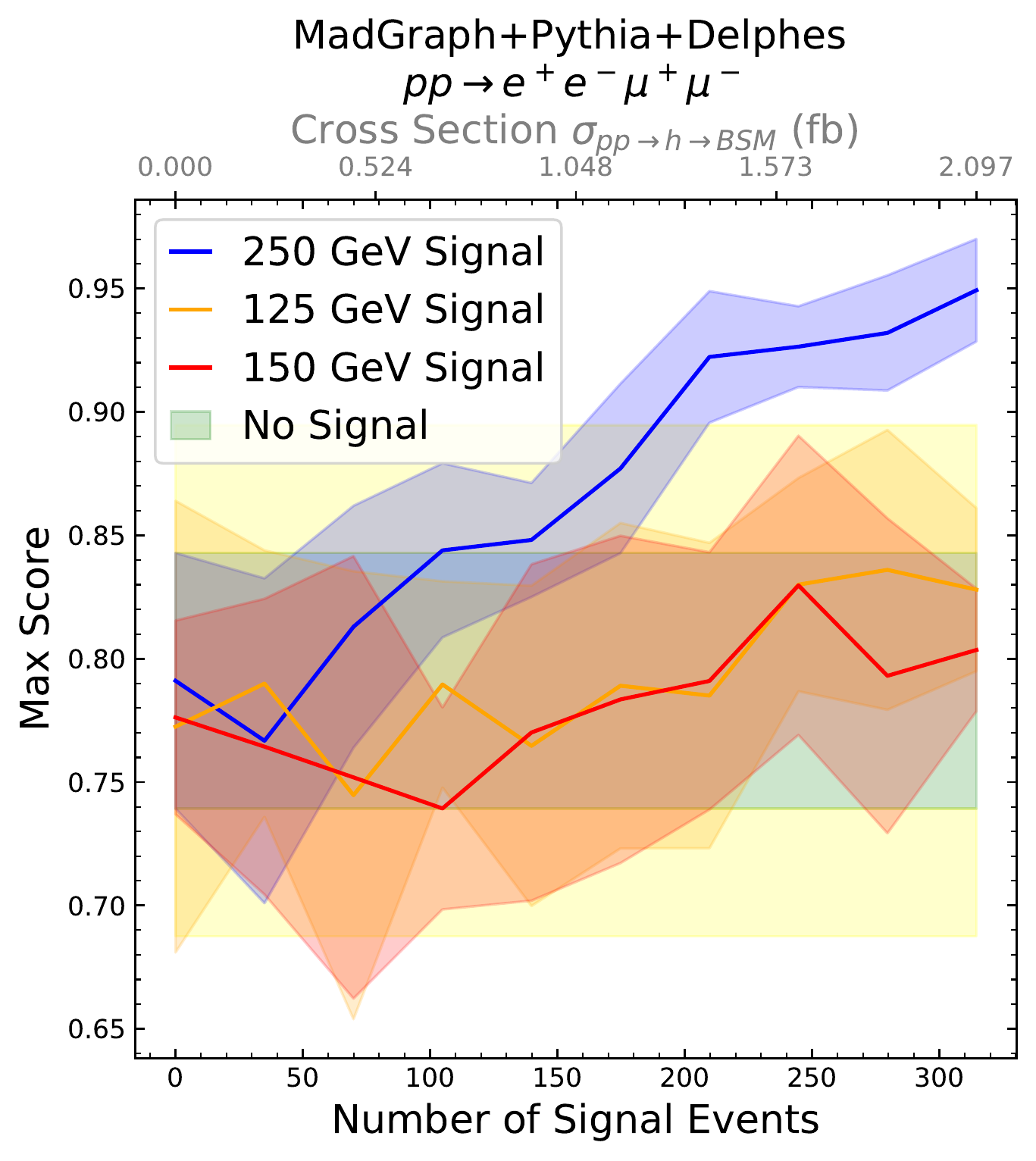}
    \caption{\textbf{BCE Loss}.  Same as Fig.~\ref{fig:BCE_zoomout}, but zoomed in by a factor of 10 on the horizontal axis.  The 95\% exclusion limits approximately correspond to where the lines for each signal cross the yellow $2\sigma$ band for the background-only hypothesis.}
    \label{fig:BCE_zoomin}
\end{figure}

The values of the four statistics described in the previous section as a function of the number of injected signal events are presented in Fig.~\ref{fig:MLC_zoomout} and~\ref{fig:MLC_zoomin} for the MLC loss and Fig.~\ref{fig:BCE_zoomout} and~\ref{fig:BCE_zoomin} for the BCE loss.  As expected, the loss decreases and the the mean/max/standard deviation of the scores increases with more signal events injected.  As the number of injected signal events goes to zero, the standard deviation over the scores goes to zero.  In the MLC case, the neural network approximates the log likelihood ratio and so the average score approaches 0 as the number of injected signal events goes to zero.  The MLC loss is approximately minus two times a $\chi^2$ random variable~\cite{DAgnolo:2018cun,2101.07263}.  In the BCE case, the neural network approximates $p(\text{`data'}|m_{e^+e^-},m_{\mu^+\mu^-},m_{4\ell})$, which approaches a $\delta$-function centered at 0.5 as the number of signal events goes to zero.  The BCE loss itself approaches $\log(2)\approx0.69$.  

The green (yellow) band corresponds to the $1\sigma$ (2$\sigma$) region of the background-only hypothesis.  The bands for each signal correspond to the standard deviation across 10 independent signal injections.  We show zoomed-out (Fig.~\ref{fig:MLC_zoomout} and~\ref{fig:BCE_zoomout}) and zoomed-in (Fig.~\ref{fig:MLC_zoomin} and~\ref{fig:BCE_zoomin}) versions to emphasize where the bands cross the $2\sigma$ line, which approximately corresponds to the 95\% exclusion limit.  For both loss functions, we find that the standard deviations over the scores is very effective.  The loss in the case of the MLC is more useful than the BCE loss and has comparable performance to the score standard deviation. This may be expected, since the MLC loss asymptotically is monotonically related to the likelihood ratio of the two samples (and thus optimal).  The max score is must more useful for BCE than for MLC.

The actual limits are comparable for both losses.  For example, the limits for the 250 GeV signal are about $0.4^{+0.1}_{-0.3}$ fb ($0.4\pm 0.2$) for MLC (BCE) and for the other signals are about $0.8^{+0.7}_{-0.2}$ fb ($1.2\pm 0.2$) for MLC (BCE).  To put these limits in context, the lmits in Ref.~\cite{ATLAS:2021ldb} (CMS is similar) for $h\rightarrow XX$ has a limit of about 0.04 fb in the $e^+e^-\mu^+\mu^-$ channel (nearly independent of $m_X$).  This limit is about 10 times better than what we find which is the price we pay for being model agnostic\footnote{In addition to the impact of the simplifying assumptions we make in the analysis presented here compared to the more detailed ATLAS/CMS searches.}.  On the other hand, the ATLAS/CMS searches have no sensitivity outside of the fixed $m_X$, $m_h$ mass windows, while we have broad sensitivity that is nearly independent of all masses.

\clearpage

\section{Conclusions}
\label{sec:concl}

In this paper, we have explored the use of less-than-supervised anomaly detection for the four-lepton final state at the LHC.  The motivation of this approach is that it is broadly sensitive to BSM models without specifying model parameters and is thus complementary to model-specific analyses.  The core methodology we use is not new.  The idea of performing AD by comparing data with a reference sample was first introduced in Ref.~\cite{Collins:2018epr,Collins:2019jip} and studied in the context of a simulated reference sample in Ref.~\cite{DAgnolo:2018cun,DAgnolo:2019vbw,dAgnolo:2021aun}.  Our contribution is to note that the four-lepton final state is special in that it is nearly unique at the LHC as a case where simulations are used directly to estimate Standard Model backgrounds.  This allows for the data-versus-simulation strategy and we have explicitly demonstrated how the method works using multiple signal models.  We have extended previous results by exploring different statistics that can be used to characterize the effective goodness of fit from the machine learning classifiers.  While we have focused on resonant anomalies, we note that the methodology studied here will also work in the case of non-resonant modifications to the four-lepton final state.  Many parts of the parameter space are statistics limited, but a full investigation of systematic uncertainties will be required for a complete experimental implementation in the future.  A variety of uncertainty- and inference-aware methods have been proposed that can be used in this case~\cite{dAgnolo:2021aun,Ghosh:2021roe,deCastro:2018mgh,Wunsch:2020iuh,Elwood:2018qsr,Xia:2018kgd,PhysRevD.97.083004,Alsing:2019dvb,Simpson:2022suz}.

With no significant evidence for physics beyond the Standard Model at the LHC so far, it is critical to extend our analysis methodologies to include more model agnostic approaches.  It is likely that many techniques will be required to achieve broad sensitivity.  We have explored a simulation-driven method in this paper for the four-lepton final state because of the existence of precise background models.  Combining simulation-based AD with data-driven background estimation strategies may be able extend the scope of similar methods to a wide variety of final states at the LHC and beyond.

\section*{\label{sec::acknowledgments}Acknowledgments}

We thank Gregor Kasieczka, Maurizio Pierini, and David Shih for useful feedback on the manuscript.  This work was supported by the Department of Energy, Office of Science under contract number DE-AC02-05CH11231.  KK was supported in part by NSF PHY REU Grant 1949923.

\bibliographystyle{JHEP}
\bibliography{main,HEPML}

\providecommand{\href}[2]{#2}\begingroup\raggedright\begin{thebibliography}{100}

\bibitem{atlasexoticstwiki}
{ATLAS Collaboration}, ``{Exotic Physics Searches}.''
  \url{https://twiki.cern.ch/twiki/bin/view/AtlasPublic/ExoticsPublicResults},
  2022.

\bibitem{atlassusytwiki}
{ATLAS Collaboration}, ``{Supersymmetry searches}.''
  \url{https://twiki.cern.ch/twiki/bin/view/AtlasPublic/SupersymmetryPublicResults},
  2022.

\bibitem{atlashdbspublictwiki}
{ATLAS Collaboration}, ``{Higgs and Diboson Searches}.''
  \url{https://twiki.cern.ch/twiki/bin/view/AtlasPublic/HDBSPublicResults},
  2022.

\bibitem{cmsexoticstwiki}
{CMS Collaboration}, ``{CMS Exotica Public Physics Results}.''
  \url{https://twiki.cern.ch/twiki/bin/view/CMSPublic/PhysicsResultsEXO}, 2022.

\bibitem{cmssusytwiki}
{CMS Collaboration}, ``{CMS Supersymmetry Physics Results}.''
  \url{https://twiki.cern.ch/twiki/bin/view/CMSPublic/PhysicsResultsSUS}, 2022.

\bibitem{cmsb2gtwiki}
{CMS Collaboration}, ``{CMS Beyond-two-generations (B2G) Public Physics
  Results}.''
  \url{https://twiki.cern.ch/twiki/bin/view/CMSPublic/PhysicsResultsB2G}, 2022.

\bibitem{lhcbtwiki}
{LHCb Collaboration}, ``{Publications of the QCD, Electroweak and Exotica
  Working Group}.''
  \url{http://lhcbproject.web.cern.ch/lhcbproject/Publications/LHCbProjectPublic/Summary_QEE.html},
  2022.

\bibitem{sleuth}
B.~Knuteson, ``{A Quasi-Model-Independent Search for New High $p_T$ Physics at
  D0}.''
  \url{https://www-d0.fnal.gov/results/publications_talks/thesis/knuteson/thesis.ps}.

\bibitem{Abbott:2000fb}
{\scshape D0} collaboration, B.~Abbott et~al., \emph{{Search for new physics in
  $e\mu X$ data at D\O\ using Sherlock: A quasi model independent search
  strategy for new physics}},
  \href{https://doi.org/10.1103/PhysRevD.62.092004}{\emph{Phys. Rev.}
  {\bfseries D62} (2000) 092004},
  [\href{https://arxiv.org/abs/hep-ex/0006011}{{\ttfamily hep-ex/0006011}}].

\bibitem{Abbott:2000gx}
{\scshape D0} collaboration, V.~M. Abazov et~al., \emph{{A Quasi model
  independent search for new physics at large transverse momentum}},
  \href{https://doi.org/10.1103/PhysRevD.64.012004}{\emph{Phys. Rev.}
  {\bfseries D64} (2001) 012004},
  [\href{https://arxiv.org/abs/hep-ex/0011067}{{\ttfamily hep-ex/0011067}}].

\bibitem{Abbott:2001ke}
{\scshape D0} collaboration, B.~Abbott et~al., \emph{{A quasi-model-independent
  search for new high $p_T$ physics at D\O}},
  \href{https://doi.org/10.1103/PhysRevLett.86.3712}{\emph{Phys. Rev. Lett.}
  {\bfseries 86} (2001) 3712--3717},
  [\href{https://arxiv.org/abs/hep-ex/0011071}{{\ttfamily hep-ex/0011071}}].

\bibitem{Aaron:2008aa}
{\scshape H1} collaboration, F.~D. Aaron et~al., \emph{{A General Search for
  New Phenomena at HERA}},
  \href{https://doi.org/10.1016/j.physletb.2009.03.034}{\emph{Phys. Lett.}
  {\bfseries B674} (2009) 257--268},
  [\href{https://arxiv.org/abs/0901.0507}{{\ttfamily 0901.0507}}].

\bibitem{Aktas:2004pz}
{\scshape H1} collaboration, A.~Aktas et~al., \emph{{A General search for new
  phenomena in ep scattering at HERA}},
  \href{https://doi.org/10.1016/j.physletb.2004.09.057}{\emph{Phys. Lett.}
  {\bfseries B602} (2004) 14--30},
  [\href{https://arxiv.org/abs/hep-ex/0408044}{{\ttfamily hep-ex/0408044}}].

\bibitem{Cranmer:2005zn}
K.~S. Cranmer, ``{Searching for new physics: Contributions to LEP and the
  LHC}.'' \url{https://inspirehep.net/literature/689610}, 2005.

\bibitem{Aaltonen:2007dg}
{\scshape CDF} collaboration, T.~Aaltonen et~al., \emph{{Model-Independent and
  Quasi-Model-Independent Search for New Physics at CDF}},
  \href{https://doi.org/10.1103/PhysRevD.78.012002}{\emph{Phys. Rev.}
  {\bfseries D78} (2008) 012002},
  [\href{https://arxiv.org/abs/0712.1311}{{\ttfamily 0712.1311}}].

\bibitem{Aaltonen:2007ab}
{\scshape CDF} collaboration, T.~Aaltonen et~al., \emph{{Model-Independent
  Global Search for New High-p(T) Physics at CDF}},
  \href{https://arxiv.org/abs/0712.2534}{{\ttfamily 0712.2534}}.

\bibitem{Aaltonen:2008vt}
{\scshape CDF} collaboration, T.~Aaltonen et~al., \emph{{Global Search for New
  Physics with 2.0 fb$^{-1}$ at CDF}},
  \href{https://doi.org/10.1103/PhysRevD.79.011101}{\emph{Phys. Rev.}
  {\bfseries D79} (2009) 011101},
  [\href{https://arxiv.org/abs/0809.3781}{{\ttfamily 0809.3781}}].

\bibitem{CMS-PAS-EXO-14-016}
{CMS Collaboration}, \emph{{MUSiC, a Model Unspecific Search for New Physics,
  in pp Collisions at $\sqrt{s}=8$ TeV}}, {\emph{CMS-PAS-EXO-14-016} (2017) }.

\bibitem{CMS-PAS-EXO-10-021}
{CMS Collaboration}, \emph{{Model Unspecific Search for New Physics in pp
  Collisions at $\sqrt{s} = 7$ TeV}}, {\emph{CMS-PAS-EXO-10-021} (2011) }.

\bibitem{CMS:2020ohc}
{CMS Collaboration}, ``{MUSiC, a model unspecific search for new physics, in
  $pp$ collisions at $\sqrt{s}=13$ TeV}.''
  \url{https://cds.cern.ch/record/2718811}, 5, 2020.

\bibitem{Sirunyan:2020jwk}
{\scshape CMS} collaboration, A.~M. Sirunyan et~al., \emph{{MUSiC: a
  model-unspecific search for new physics in proton\textendash{}proton
  collisions at $\sqrt{s} = 13\,\text {TeV} $}},
  \href{https://doi.org/10.1140/epjc/s10052-021-09236-z}{\emph{Eur. Phys. J. C}
  {\bfseries 81} (2021) 629},
  [\href{https://arxiv.org/abs/2010.02984}{{\ttfamily 2010.02984}}].

\bibitem{Aaboud:2018ufy}
{\scshape ATLAS} collaboration, M.~Aaboud et~al., \emph{{A strategy for a
  general search for new phenomena using data-derived signal regions and its
  application within the ATLAS experiment}},
  \href{https://doi.org/10.1140/epjc/s10052-019-6540-y}{\emph{Eur. Phys. J.}
  {\bfseries C79} (2019) 120},
  [\href{https://arxiv.org/abs/1807.07447}{{\ttfamily 1807.07447}}].

\bibitem{ATLAS-CONF-2014-006}
{\scshape {ATLAS}} collaboration, \emph{{A general search for new phenomena
  with the ATLAS detector in pp collisions at $\sqrt{s}=8$ TeV}},
  {\emph{ATLAS-CONF-2014-006} (Mar, 2014) }.

\bibitem{ATLAS-CONF-2012-107}
{\scshape ATLAS} collaboration, \emph{{A general search for new phenomena with
  the ATLAS detector in pp collisions at $\sqrt{s}=7$ TeV}},
  {\emph{ATLAS-CONF-2012-107} (2012) }.

\bibitem{DAgnolo:2018cun}
R.~T. D'Agnolo and A.~Wulzer, \emph{{Learning New Physics from a Machine}},
  \href{https://doi.org/10.1103/PhysRevD.99.015014}{\emph{Phys. Rev.}
  {\bfseries D99} (2019) 015014},
  [\href{https://arxiv.org/abs/1806.02350}{{\ttfamily 1806.02350}}].

\bibitem{Collins:2018epr}
J.~H. Collins, K.~Howe and B.~Nachman, \emph{{Anomaly Detection for Resonant
  New Physics with Machine Learning}},
  \href{https://doi.org/10.1103/PhysRevLett.121.241803}{\emph{Phys. Rev. Lett.}
  {\bfseries 121} (2018) 241803},
  [\href{https://arxiv.org/abs/1805.02664}{{\ttfamily 1805.02664}}].

\bibitem{Collins:2019jip}
J.~H. Collins, K.~Howe and B.~Nachman, \emph{{Extending the search for new
  resonances with machine learning}},
  \href{https://doi.org/10.1103/PhysRevD.99.014038}{\emph{Phys. Rev.}
  {\bfseries D99} (2019) 014038},
  [\href{https://arxiv.org/abs/1902.02634}{{\ttfamily 1902.02634}}].

\bibitem{DAgnolo:2019vbw}
R.~T. D'Agnolo, G.~Grosso, M.~Pierini, A.~Wulzer and M.~Zanetti,
  \emph{{Learning multivariate new physics}},
  \href{https://doi.org/10.1140/epjc/s10052-021-08853-y}{\emph{Eur. Phys. J. C}
  {\bfseries 81} (2021) 89},
  [\href{https://arxiv.org/abs/1912.12155}{{\ttfamily 1912.12155}}].

\bibitem{Andreassen:2020nkr}
A.~Andreassen, B.~Nachman and D.~Shih, \emph{{Simulation Assisted
  Likelihood-free Anomaly Detection}},
  \href{https://doi.org/10.1103/PhysRevD.101.095004}{\emph{Phys. Rev. D}
  {\bfseries 101} (2020) 095004},
  [\href{https://arxiv.org/abs/2001.05001}{{\ttfamily 2001.05001}}].

\bibitem{Nachman:2020lpy}
B.~Nachman and D.~Shih, \emph{{Anomaly Detection with Density Estimation}},
  \href{https://doi.org/10.1103/PhysRevD.101.075042}{\emph{Phys. Rev. D}
  {\bfseries 101} (2020) 075042},
  [\href{https://arxiv.org/abs/2001.04990}{{\ttfamily 2001.04990}}].

\bibitem{Hallin:2021wme}
A.~Hallin, J.~Isaacson, G.~Kasieczka, C.~Krause, B.~Nachman, T.~Quadfasel
  et~al., \emph{{Classifying Anomalies THrough Outer Density Estimation
  (CATHODE)}},  \href{https://arxiv.org/abs/2109.00546}{{\ttfamily
  2109.00546}}.

\bibitem{Farina:2018fyg}
M.~Farina, Y.~Nakai and D.~Shih, \emph{{Searching for New Physics with Deep
  Autoencoders}},
  \href{https://doi.org/10.1103/PhysRevD.101.075021}{\emph{Phys. Rev. D}
  {\bfseries 101} (2020) 075021},
  [\href{https://arxiv.org/abs/1808.08992}{{\ttfamily 1808.08992}}].

\bibitem{Heimel:2018mkt}
T.~Heimel, G.~Kasieczka, T.~Plehn and J.~M. Thompson, \emph{{QCD or What?}},
  \href{https://doi.org/10.21468/SciPostPhys.6.3.030}{\emph{SciPost Phys.}
  {\bfseries 6} (2019) 030},
  [\href{https://arxiv.org/abs/1808.08979}{{\ttfamily 1808.08979}}].

\bibitem{Roy:2019jae}
T.~S. Roy and A.~H. Vijay, \emph{{A robust anomaly finder based on
  autoencoder}},  \href{https://arxiv.org/abs/1903.02032}{{\ttfamily
  1903.02032}}.

\bibitem{Cerri:2018anq}
O.~Cerri, T.~Q. Nguyen, M.~Pierini, M.~Spiropulu and J.-R. Vlimant,
  \emph{{Variational Autoencoders for New Physics Mining at the Large Hadron
  Collider}}, \href{https://doi.org/10.1007/JHEP05(2019)036}{\emph{JHEP}
  {\bfseries 05} (2019) 036},
  [\href{https://arxiv.org/abs/1811.10276}{{\ttfamily 1811.10276}}].

\bibitem{Blance:2019ibf}
A.~Blance, M.~Spannowsky and P.~Waite, \emph{{Adversarially-trained
  autoencoders for robust unsupervised new physics searches}},
  \href{https://doi.org/10.1007/JHEP10(2019)047}{\emph{JHEP} {\bfseries 10}
  (2019) 047}, [\href{https://arxiv.org/abs/1905.10384}{{\ttfamily
  1905.10384}}].

\bibitem{Hajer:2018kqm}
J.~Hajer, Y.-Y. Li, T.~Liu and H.~Wang, \emph{{Novelty Detection Meets Collider
  Physics}}, \href{https://doi.org/10.1103/PhysRevD.101.076015}{\emph{Phys.
  Rev. D} {\bfseries 101} (2020) 076015},
  [\href{https://arxiv.org/abs/1807.10261}{{\ttfamily 1807.10261}}].

\bibitem{DeSimone:2018efk}
A.~De~Simone and T.~Jacques, \emph{{Guiding New Physics Searches with
  Unsupervised Learning}},
  \href{https://doi.org/10.1140/epjc/s10052-019-6787-3}{\emph{Eur. Phys. J.}
  {\bfseries C79} (2019) 289},
  [\href{https://arxiv.org/abs/1807.06038}{{\ttfamily 1807.06038}}].

\bibitem{Mullin:2019mmh}
A.~Mullin, S.~Nicholls, H.~Pacey, M.~Parker, M.~White and S.~Williams,
  \emph{{Does SUSY have friends? A new approach for LHC event analysis}},
  \href{https://doi.org/10.1007/JHEP02(2021)160}{\emph{JHEP} {\bfseries 02}
  (2021) 160}, [\href{https://arxiv.org/abs/1912.10625}{{\ttfamily
  1912.10625}}].

\bibitem{1809.02977}
G.~M. Alessandro~Casa, \emph{{Nonparametric semisupervised classification for
  signal detection in high energy physics}},
  \href{https://arxiv.org/abs/1809.02977}{{\ttfamily 1809.02977}}.

\bibitem{Dillon:2019cqt}
B.~M. Dillon, D.~A. Faroughy and J.~F. Kamenik, \emph{{Uncovering latent jet
  substructure}},
  \href{https://doi.org/10.1103/PhysRevD.100.056002}{\emph{Phys. Rev.}
  {\bfseries D100} (2019) 056002},
  [\href{https://arxiv.org/abs/1904.04200}{{\ttfamily 1904.04200}}].

\bibitem{Aguilar-Saavedra:2017rzt}
J.~A. Aguilar-Saavedra, J.~H. Collins and R.~K. Mishra, \emph{{A generic
  anti-QCD jet tagger}},
  \href{https://doi.org/10.1007/JHEP11(2017)163}{\emph{JHEP} {\bfseries 11}
  (2017) 163}, [\href{https://arxiv.org/abs/1709.01087}{{\ttfamily
  1709.01087}}].

\bibitem{Romao:2019dvs}
M.~Romão~Crispim, N.~Castro, R.~Pedro and T.~Vale, \emph{{Transferability of
  Deep Learning Models in Searches for New Physics at Colliders}},
  \href{https://doi.org/10.1103/PhysRevD.101.035042}{\emph{Phys.\ Rev.\ D}
  {\bfseries 101} (2020) 035042},
  [\href{https://arxiv.org/abs/1912.04220}{{\ttfamily 1912.04220}}].

\bibitem{Romao:2020ojy}
M.~Crispim Rom\~ao, N.~F. Castro, J.~G. Milhano, R.~Pedro and T.~Vale,
  \emph{{Use of a generalized energy Mover\textquoteright{}s distance in the
  search for rare phenomena at colliders}},
  \href{https://doi.org/10.1140/epjc/s10052-021-08891-6}{\emph{Eur. Phys. J. C}
  {\bfseries 81} (2021) 192},
  [\href{https://arxiv.org/abs/2004.09360}{{\ttfamily 2004.09360}}].

\bibitem{knapp2020adversarially}
O.~Knapp, O.~Cerri, G.~Dissertori, T.~Q. Nguyen, M.~Pierini and J.-R. Vlimant,
  \emph{{Adversarially Learned Anomaly Detection on CMS Open Data:
  re-discovering the top quark}},
  \href{https://doi.org/10.1140/epjp/s13360-021-01109-4}{\emph{Eur. Phys. J.
  Plus} {\bfseries 136} (2021) 236},
  [\href{https://arxiv.org/abs/2005.01598}{{\ttfamily 2005.01598}}].

\bibitem{collaboration2020dijet}
{\scshape ATLAS} collaboration, G.~Aad et~al., \emph{{Dijet resonance search
  with weak supervision using $\sqrt{s}=13$ TeV $pp$ collisions in the ATLAS
  detector}}, \href{https://doi.org/10.1103/PhysRevLett.125.131801}{\emph{Phys.
  Rev. Lett.} {\bfseries 125} (2020) 131801},
  [\href{https://arxiv.org/abs/2005.02983}{{\ttfamily 2005.02983}}].

\bibitem{1797846}
B.~M. Dillon, D.~A. Faroughy, J.~F. Kamenik and M.~Szewc, \emph{{Learning the
  latent structure of collider events}},
  \href{https://arxiv.org/abs/2005.12319}{{\ttfamily 2005.12319}}.

\bibitem{1800445}
M.~Crispim Rom\~ao, N.~F. Castro and R.~Pedro, \emph{{Finding New Physics
  without learning about it: Anomaly Detection as a tool for Searches at
  Colliders}},
  \href{https://doi.org/10.1140/epjc/s10052-021-09813-2}{\emph{Eur. Phys. J. C}
  {\bfseries 81} (2021) 27},
  [\href{https://arxiv.org/abs/2006.05432}{{\ttfamily 2006.05432}}].

\bibitem{Amram:2020ykb}
O.~Amram and C.~M. Suarez, \emph{{Tag N\textquoteright{} Train: a technique to
  train improved classifiers on unlabeled data}},
  \href{https://doi.org/10.1007/JHEP01(2021)153}{\emph{JHEP} {\bfseries 01}
  (2021) 153}, [\href{https://arxiv.org/abs/2002.12376}{{\ttfamily
  2002.12376}}].

\bibitem{Cheng:2020dal}
T.~Cheng, J.-F. Arguin, J.~Leissner-Martin, J.~Pilette and T.~Golling,
  \emph{{Variational Autoencoders for Anomalous Jet Tagging}},
  \href{https://arxiv.org/abs/2007.01850}{{\ttfamily 2007.01850}}.

\bibitem{Khosa:2020qrz}
C.~K. Khosa and V.~Sanz, \emph{{Anomaly Awareness}},
  \href{https://arxiv.org/abs/2007.14462}{{\ttfamily 2007.14462}}.

\bibitem{Thaprasop:2020mzp}
P.~Thaprasop, K.~Zhou, J.~Steinheimer and C.~Herold, \emph{{Unsupervised
  Outlier Detection in Heavy-Ion Collisions}},
  \href{https://arxiv.org/abs/2007.15830}{{\ttfamily 2007.15830}}.

\bibitem{Alexander:2020mbx}
S.~Alexander, S.~Gleyzer, H.~Parul, P.~Reddy, M.~W. Toomey, E.~Usai et~al.,
  \emph{{Decoding Dark Matter Substructure without Supervision}},
  \href{https://arxiv.org/abs/2008.12731}{{\ttfamily 2008.12731}}.

\bibitem{aguilarsaavedra2020mass}
J.~A. Aguilar-Saavedra, F.~R. Joaquim and J.~F. Seabra, \emph{{Mass Unspecific
  Supervised Tagging (MUST) for boosted jets}},
  \href{https://doi.org/10.1007/JHEP03(2021)012}{\emph{JHEP} {\bfseries 03}
  (2021) 012}, [\href{https://arxiv.org/abs/2008.12792}{{\ttfamily
  2008.12792}}].

\bibitem{1815227}
K.~Benkendorfer, L.~L. Pottier and B.~Nachman, \emph{{Simulation-assisted
  decorrelation for resonant anomaly detection}},
  \href{https://doi.org/10.1103/PhysRevD.104.035003}{\emph{Phys. Rev. D}
  {\bfseries 104} (2021) 035003},
  [\href{https://arxiv.org/abs/2009.02205}{{\ttfamily 2009.02205}}].

\bibitem{pol2020anomaly}
{Adrian Alan Pol and Victor Berger and Gianluca Cerminara and Cecile Germain
  and Maurizio Pierini}, \emph{{Anomaly Detection With Conditional Variational
  Autoencoders}},  \href{https://arxiv.org/abs/2010.05531}{{\ttfamily
  2010.05531}}.

\bibitem{Mikuni:2020qds}
V.~Mikuni and F.~Canelli, \emph{{Unsupervised clustering for collider
  physics}},  \href{https://arxiv.org/abs/2010.07106}{{\ttfamily 2010.07106}}.

\bibitem{vanBeekveld:2020txa}
M.~van Beekveld, S.~Caron, L.~Hendriks, P.~Jackson, A.~Leinweber, S.~Otten
  et~al., \emph{{Combining outlier analysis algorithms to identify new physics
  at the LHC}}, \href{https://doi.org/10.1007/JHEP09(2021)024}{\emph{JHEP}
  {\bfseries 09} (2021) 024},
  [\href{https://arxiv.org/abs/2010.07940}{{\ttfamily 2010.07940}}].

\bibitem{Park:2020pak}
S.~E. Park, D.~Rankin, S.-M. Udrescu, M.~Yunus and P.~Harris, \emph{{Quasi
  Anomalous Knowledge: Searching for new physics with embedded knowledge}},
  \href{https://doi.org/10.1007/JHEP06(2021)030}{\emph{JHEP} {\bfseries 21}
  (2020) 030}, [\href{https://arxiv.org/abs/2011.03550}{{\ttfamily
  2011.03550}}].

\bibitem{Faroughy:2020gas}
D.~A. Faroughy, \emph{{Uncovering hidden patterns in collider events with
  Bayesian probabilistic models}},
  \href{https://arxiv.org/abs/2012.08579}{{\ttfamily 2012.08579}}.

\bibitem{Stein:2020rou}
G.~Stein, U.~Seljak and B.~Dai, \emph{{Unsupervised in-distribution anomaly
  detection of new physics through conditional density estimation}},
  \href{https://arxiv.org/abs/2012.11638}{{\ttfamily 2012.11638}}.

\bibitem{Chakravarti:2021svb}
P.~Chakravarti, M.~Kuusela, J.~Lei and L.~Wasserman, \emph{{Model-Independent
  Detection of New Physics Signals Using Interpretable Semi-Supervised
  Classifier Tests}},  \href{https://arxiv.org/abs/2102.07679}{{\ttfamily
  2102.07679}}.

\bibitem{Batson:2021agz}
J.~Batson, C.~G. Haaf, Y.~Kahn and D.~A. Roberts, \emph{{Topological
  Obstructions to Autoencoding}},
  \href{https://doi.org/10.1007/JHEP04(2021)280}{\emph{JHEP} {\bfseries 04}
  (2021) 280}, [\href{https://arxiv.org/abs/2102.08380}{{\ttfamily
  2102.08380}}].

\bibitem{Blance:2021gcs}
A.~Blance and M.~Spannowsky, \emph{{Unsupervised event classification with
  graphs on classical and photonic quantum computers}},
  \href{https://doi.org/10.1007/JHEP08(2021)170}{\emph{JHEP} {\bfseries 21}
  (2020) 170}, [\href{https://arxiv.org/abs/2103.03897}{{\ttfamily
  2103.03897}}].

\bibitem{Bortolato:2021zic}
B.~Bortolato, B.~M. Dillon, J.~F. Kamenik and A.~Smolkovi\v{c}, \emph{{Bump
  Hunting in Latent Space}},
  \href{https://arxiv.org/abs/2103.06595}{{\ttfamily 2103.06595}}.

\bibitem{Collins:2021nxn}
J.~H. Collins, P.~Mart\'\i{}n-Ramiro, B.~Nachman and D.~Shih, \emph{{Comparing
  weak- and unsupervised methods for resonant anomaly detection}},
  \href{https://doi.org/10.1140/epjc/s10052-021-09389-x}{\emph{Eur. Phys. J. C}
  {\bfseries 81} (2021) 617},
  [\href{https://arxiv.org/abs/2104.02092}{{\ttfamily 2104.02092}}].

\bibitem{Dillon:2021nxw}
B.~M. Dillon, T.~Plehn, C.~Sauer and P.~Sorrenson, \emph{{Better Latent Spaces
  for Better Autoencoders}},
  \href{https://doi.org/10.21468/SciPostPhys.11.3.061}{\emph{SciPost Phys.}
  {\bfseries 11} (2021) 061},
  [\href{https://arxiv.org/abs/2104.08291}{{\ttfamily 2104.08291}}].

\bibitem{Finke:2021sdf}
T.~Finke, M.~Kr\"amer, A.~Morandini, A.~M\"uck and I.~Oleksiyuk,
  \emph{{Autoencoders for unsupervised anomaly detection in high energy
  physics}}, \href{https://doi.org/10.1007/JHEP06(2021)161}{\emph{JHEP}
  {\bfseries 06} (2021) 161},
  [\href{https://arxiv.org/abs/2104.09051}{{\ttfamily 2104.09051}}].

\bibitem{Shih:2021kbt}
D.~Shih, M.~R. Buckley, L.~Necib and J.~Tamanas, \emph{{via machinae: Searching
  for stellar streams using unsupervised machine learning}},
  \href{https://doi.org/10.1093/mnras/stab3372}{\emph{Mon. Not. Roy. Astron.
  Soc.} {\bfseries 509} (2021) 5992--6007},
  [\href{https://arxiv.org/abs/2104.12789}{{\ttfamily 2104.12789}}].

\bibitem{Atkinson:2021nlt}
O.~Atkinson, A.~Bhardwaj, C.~Englert, V.~S. Ngairangbam and M.~Spannowsky,
  \emph{{Anomaly detection with convolutional Graph Neural Networks}},
  \href{https://doi.org/10.1007/JHEP08(2021)080}{\emph{JHEP} {\bfseries 08}
  (2021) 080}, [\href{https://arxiv.org/abs/2105.07988}{{\ttfamily
  2105.07988}}].

\bibitem{Kahn:2021drv}
A.~Kahn, J.~Gonski, I.~Ochoa, D.~Williams and G.~Brooijmans, \emph{{Anomalous
  jet identification via sequence modeling}},
  \href{https://doi.org/10.1088/1748-0221/16/08/P08012}{\emph{JINST} {\bfseries
  16} (2021) P08012}, [\href{https://arxiv.org/abs/2105.09274}{{\ttfamily
  2105.09274}}].

\bibitem{Dorigo:2021iyy}
T.~Dorigo, M.~Fumanelli, C.~Maccani, M.~Mojsovska, G.~C. Strong and B.~Scarpa,
  \emph{{RanBox: Anomaly Detection in the Copula Space}},
  \href{https://arxiv.org/abs/2106.05747}{{\ttfamily 2106.05747}}.

\bibitem{Caron:2021wmq}
S.~Caron, L.~Hendriks and R.~Verheyen, \emph{{Rare and Different: Anomaly
  Scores from a combination of likelihood and out-of-distribution models to
  detect new physics at the LHC}},
  \href{https://doi.org/10.21468/SciPostPhys.12.2.077}{\emph{SciPost Phys.}
  {\bfseries 12} (2022) 077},
  [\href{https://arxiv.org/abs/2106.10164}{{\ttfamily 2106.10164}}].

\bibitem{Govorkova:2021hqu}
E.~Govorkova, E.~Puljak, T.~Aarrestad, M.~Pierini, K.~A. Wo\'zniak and
  J.~Ngadiuba, \emph{{LHC physics dataset for unsupervised New Physics
  detection at 40 MHz}},  \href{https://arxiv.org/abs/2107.02157}{{\ttfamily
  2107.02157}}.

\bibitem{Kasieczka:2021tew}
G.~Kasieczka, B.~Nachman and D.~Shih, \emph{{New Methods and Datasets for Group
  Anomaly Detection From Fundamental Physics}},  7, 2021,
  \href{https://arxiv.org/abs/2107.02821}{{\ttfamily 2107.02821}}.

\bibitem{Volkovich:2021txe}
S.~Volkovich, F.~D.~V. Halevy and S.~Bressler, \emph{{The Data-Directed
  Paradigm for BSM searches}},
  \href{https://arxiv.org/abs/2107.11573}{{\ttfamily 2107.11573}}.

\bibitem{Govorkova:2021utb}
E.~Govorkova et~al., \emph{{Autoencoders on FPGAs for real-time, unsupervised
  new physics detection at 40 MHz at the Large Hadron Collider}},
  \href{https://arxiv.org/abs/2108.03986}{{\ttfamily 2108.03986}}.

\bibitem{Ostdiek:2021bem}
B.~Ostdiek, \emph{{Deep Set Auto Encoders for Anomaly Detection in Particle
  Physics}},  \href{https://arxiv.org/abs/2109.01695}{{\ttfamily 2109.01695}}.

\bibitem{Fraser:2021lxm}
K.~Fraser, S.~Homiller, R.~K. Mishra, B.~Ostdiek and M.~D. Schwartz,
  \emph{{Challenges for Unsupervised Anomaly Detection in Particle Physics}},
  \href{https://arxiv.org/abs/2110.06948}{{\ttfamily 2110.06948}}.

\bibitem{Kasieczka:2021xcg}
G.~Kasieczka et~al., \emph{{The LHC Olympics 2020: A Community Challenge for
  Anomaly Detection in High Energy Physics}},
  \href{https://arxiv.org/abs/2101.08320}{{\ttfamily 2101.08320}}.

\bibitem{Aarrestad:2021oeb}
T.~Aarrestad et~al., \emph{{The Dark Machines Anomaly Score Challenge:
  Benchmark Data and Model Independent Event Classification for the Large
  Hadron Collider}},
  \href{https://doi.org/10.21468/SciPostPhys.12.1.043}{\emph{SciPost Phys.}
  {\bfseries 12} (2022) 043},
  [\href{https://arxiv.org/abs/2105.14027}{{\ttfamily 2105.14027}}].

\bibitem{2112.03769}
{G. Karagiorgi, G. Kasieczka, S. Kravitz, B. Nachman, D. Shih}, \emph{{Machine
  Learning in the Search for New Fundamental Physics}},
  \href{https://arxiv.org/abs/2112.03769}{{\ttfamily 2112.03769}}.

\bibitem{Feickert:2021ajf}
M.~Feickert and B.~Nachman, \emph{{A Living Review of Machine Learning for
  Particle Physics}},  \href{https://arxiv.org/abs/2102.02770}{{\ttfamily
  2102.02770}}.

\bibitem{dAgnolo:2021aun}
R.~T. d'Agnolo, G.~Grosso, M.~Pierini, A.~Wulzer and M.~Zanetti,
  \emph{{Learning New Physics from an Imperfect Machine}},
  \href{https://arxiv.org/abs/2111.13633}{{\ttfamily 2111.13633}}.

\bibitem{ATLAS:2020tlo}
{\scshape ATLAS} collaboration, G.~Aad et~al., \emph{{Search for heavy
  resonances decaying into a pair of Z bosons in the $\ell ^+\ell ^-\ell
  '^+\ell '^-$ and $\ell ^+\ell ^-\nu {{\bar{\nu }}}$ final states using 139
  $\mathrm {fb}^{-1}$ of proton\textendash{}proton collisions at $\sqrt{s} =
  13\,$TeV with the ATLAS detector}},
  \href{https://doi.org/10.1140/epjc/s10052-021-09013-y}{\emph{Eur. Phys. J. C}
  {\bfseries 81} (2021) 332},
  [\href{https://arxiv.org/abs/2009.14791}{{\ttfamily 2009.14791}}].

\bibitem{ATLAS:2018coo}
{\scshape ATLAS} collaboration, M.~Aaboud et~al., \emph{{Search for Higgs boson
  decays to beyond-the-Standard-Model light bosons in four-lepton events with
  the ATLAS detector at $\sqrt{s}=13$ TeV}},
  \href{https://doi.org/10.1007/JHEP06(2018)166}{\emph{JHEP} {\bfseries 06}
  (2018) 166}, [\href{https://arxiv.org/abs/1802.03388}{{\ttfamily
  1802.03388}}].

\bibitem{ATLAS:2020wny}
{\scshape ATLAS} collaboration, G.~Aad et~al., \emph{{Measurements of the Higgs
  boson inclusive and differential fiducial cross sections in the 4$\ell$ decay
  channel at $\sqrt{s}$ = 13 TeV}},
  \href{https://doi.org/10.1140/epjc/s10052-020-8223-0}{\emph{Eur. Phys. J. C}
  {\bfseries 80} (2020) 942},
  [\href{https://arxiv.org/abs/2004.03969}{{\ttfamily 2004.03969}}].

\bibitem{ATLAS:2021ldb}
{\scshape ATLAS} collaboration, G.~Aad et~al., \emph{{Search for Higgs bosons
  decaying into new spin-0 or spin-1 particles in four-lepton final states with
  the ATLAS detector with 139 fb$^{-1}$ of $pp$ collision data at $\sqrt{s}=13$
  TeV}},  \href{https://arxiv.org/abs/2110.13673}{{\ttfamily 2110.13673}}.

\bibitem{CMS:2016ilx}
{\scshape CMS} collaboration, \emph{{Measurements of properties of the Higgs
  boson and search for an additional resonance in the four-lepton final state
  at $\sqrt{s} = 13$ TeV}}, {\emph{CMS-PAS-HIG-16-033} (2016) }.

\bibitem{CMS:2020bni}
{\scshape CMS} collaboration, \emph{{Search for a low-mass dilepton resonance
  in Higgs boson decays to four-lepton final states at
  $\sqrt{s}=13~\mathrm{TeV}$}}, {\emph{CMS-PAS-HIG-19-007} (2020) }.

\bibitem{cmscollaboration2021search}
{CMS collaboration}, \emph{{Search for low-mass dilepton resonances in Higgs
  boson decays to four-lepton final states in proton-proton collisions at
  $\sqrt{s}$ =13 TeV}},  \href{https://arxiv.org/abs/2111.01299}{{\ttfamily
  2111.01299}}.

\bibitem{CMS:2021nnc}
{\scshape CMS} collaboration, A.~M. Sirunyan et~al., \emph{{Constraints on
  anomalous Higgs boson couplings to vector bosons and fermions in its
  production and decay using the four-lepton final state}},
  \href{https://doi.org/10.1103/PhysRevD.104.052004}{\emph{Phys. Rev. D}
  {\bfseries 104} (2021) 052004},
  [\href{https://arxiv.org/abs/2104.12152}{{\ttfamily 2104.12152}}].

\bibitem{Alwall:2014hca}
J.~Alwall, R.~Frederix, S.~Frixione, V.~Hirschi, F.~Maltoni, O.~Mattelaer
  et~al., \emph{{The automated computation of tree-level and next-to-leading
  order differential cross sections, and their matching to parton shower
  simulations}}, \href{https://doi.org/10.1007/JHEP07(2014)079}{\emph{JHEP}
  {\bfseries 07} (2014) 079},
  [\href{https://arxiv.org/abs/1405.0301}{{\ttfamily 1405.0301}}].

\bibitem{Sjostrand:2006za}
T.~Sjostrand, S.~Mrenna and P.~Z. Skands, \emph{{PYTHIA 6.4 Physics and
  Manual}}, \href{https://doi.org/10.1088/1126-6708/2006/05/026}{\emph{JHEP}
  {\bfseries 05} (2006) 026},
  [\href{https://arxiv.org/abs/hep-ph/0603175}{{\ttfamily hep-ph/0603175}}].

\bibitem{Sjostrand:2007gs}
T.~Sjostrand, S.~Mrenna and P.~Z. Skands, \emph{{A Brief Introduction to PYTHIA
  8.1}}, \href{https://doi.org/10.1016/j.cpc.2008.01.036}{\emph{Comput. Phys.
  Commun.} {\bfseries 178} (2008) 852--867},
  [\href{https://arxiv.org/abs/0710.3820}{{\ttfamily 0710.3820}}].

\bibitem{Sjostrand:2014zea}
T.~Sj\"ostrand, S.~Ask, J.~R. Christiansen, R.~Corke, N.~Desai, P.~Ilten
  et~al., \emph{{An introduction to PYTHIA 8.2}},
  \href{https://doi.org/10.1016/j.cpc.2015.01.024}{\emph{Comput. Phys. Commun.}
  {\bfseries 191} (2015) 159--177},
  [\href{https://arxiv.org/abs/1410.3012}{{\ttfamily 1410.3012}}].

\bibitem{deFavereau:2013fsa}
{\scshape DELPHES 3} collaboration, J.~de~Favereau, C.~Delaere, P.~Demin,
  A.~Giammanco, V.~Lema{\^\i}tre, A.~Mertens et~al., \emph{{DELPHES 3, A
  modular framework for fast simulation of a generic collider experiment}},
  \href{https://doi.org/10.1007/JHEP02(2014)057}{\emph{JHEP} {\bfseries 02}
  (2014) 057}, [\href{https://arxiv.org/abs/1307.6346}{{\ttfamily 1307.6346}}].

\bibitem{Mertens:2015kba}
A.~Mertens, \emph{{New features in Delphes 3}},
  \href{https://doi.org/10.1088/1742-6596/608/1/012045}{\emph{J. Phys. Conf.
  Ser.} {\bfseries 608} (2015) 012045}.

\bibitem{Selvaggi:2014mya}
M.~Selvaggi, \emph{{DELPHES 3: A modular framework for fast-simulation of
  generic collider experiments}},
  \href{https://doi.org/10.1088/1742-6596/523/1/012033}{\emph{J. Phys. Conf.
  Ser.} {\bfseries 523} (2014) 012033}.

\bibitem{Zyla:2020zbs}
{\scshape Particle Data Group} collaboration, P.~Zyla et~al., \emph{{Review of
  Particle Physics}}, \href{https://doi.org/10.1093/ptep/ptaa104}{\emph{PTEP}
  {\bfseries 2020} (2020) 083C01}.

\bibitem{efron1979}
B.~Efron, \emph{Bootstrap methods: Another look at the jackknife},
  \href{https://doi.org/10.1214/aos/1176344552}{\emph{Ann. Statist.} {\bfseries
  7} (01, 1979) 1--26}.

\bibitem{neyman1933ix}
J.~Neyman and E.~S. Pearson, \emph{On the problem of the most efficient tests
  of statistical hypotheses}, {\emph{Phil. Trans. R. Soc. Lond. A} {\bfseries
  231} (1933) 289}.

\bibitem{tensorflow}
M.~Abadi, P.~Barham, J.~Chen, Z.~Chen, A.~Davis, J.~Dean et~al.,
  \emph{Tensorflow: A system for large-scale machine learning.}, {\emph{OSDI}
  {\bfseries 16} (2016) 265}.

\bibitem{keras}
F.~Chollet, ``Keras.'' \url{https://github.com/fchollet/keras}, 2017.

\bibitem{adam}
D.~P. Kingma and J.~Ba, \emph{Adam: A method for stochastic optimization},
  \href{https://arxiv.org/abs/1412.6980}{{\ttfamily 1412.6980}}.

\bibitem{2101.07263}
B.~Nachman and J.~Thaler, \emph{{Learning from many collider events at once}},
  \href{https://doi.org/10.1103/PhysRevD.103.116013}{\emph{Phys. Rev. D}
  {\bfseries 103} (2021) 116013},
  [\href{https://arxiv.org/abs/2101.07263}{{\ttfamily 2101.07263}}].

\bibitem{Ghosh:2021roe}
A.~Ghosh, B.~Nachman and D.~Whiteson, \emph{{Uncertainty-aware machine learning
  for high energy physics}},
  \href{https://doi.org/10.1103/PhysRevD.104.056026}{\emph{Phys. Rev. D}
  {\bfseries 104} (2021) 056026},
  [\href{https://arxiv.org/abs/2105.08742}{{\ttfamily 2105.08742}}].

\bibitem{deCastro:2018mgh}
P.~De~Castro and T.~Dorigo, \emph{{INFERNO: Inference-Aware Neural
  Optimisation}},
  \href{https://doi.org/10.1016/j.cpc.2019.06.007}{\emph{Comput. Phys. Commun.}
  {\bfseries 244} (2019) 170--179},
  [\href{https://arxiv.org/abs/1806.04743}{{\ttfamily 1806.04743}}].

\bibitem{Wunsch:2020iuh}
S.~Wunsch, S.~J\"orger, R.~Wolf and G.~Quast, \emph{{Optimal Statistical
  Inference in the Presence of Systematic Uncertainties Using Neural Network
  Optimization Based on Binned Poisson Likelihoods with Nuisance Parameters}},
  \href{https://doi.org/10.1007/s41781-020-00049-5}{\emph{Comput. Softw. Big
  Sci.} {\bfseries 5} (2021) 4},
  [\href{https://arxiv.org/abs/2003.07186}{{\ttfamily 2003.07186}}].

\bibitem{Elwood:2018qsr}
A.~Elwood and D.~Kr\"ucker, \emph{{Direct optimisation of the discovery
  significance when training neural networks to search for new physics in
  particle colliders}},  \href{https://arxiv.org/abs/1806.00322}{{\ttfamily
  1806.00322}}.

\bibitem{Xia:2018kgd}
L.-G. Xia, \emph{{QBDT, a new boosting decision tree method with systematical
  uncertainties into training for High Energy Physics}},
  \href{https://doi.org/10.1016/j.nima.2019.03.088}{\emph{Nucl. Instrum. Meth.}
  {\bfseries A930} (2019) 15--26},
  [\href{https://arxiv.org/abs/1810.08387}{{\ttfamily 1810.08387}}].

\bibitem{PhysRevD.97.083004}
T.~Charnock, G.~Lavaux and B.~D. Wandelt, \emph{Automatic physical inference
  with information maximizing neural networks},
  \href{https://doi.org/10.1103/PhysRevD.97.083004}{\emph{Phys. Rev. D}
  {\bfseries 97} (Apr, 2018) 083004}.

\bibitem{Alsing:2019dvb}
J.~Alsing and B.~Wandelt, \emph{{Nuisance hardened data compression for fast
  likelihood-free inference}},
  \href{https://doi.org/10.1093/mnras/stz1900}{\emph{Mon. Not. Roy. Astron.
  Soc.} {\bfseries 488} (2019) 5093--5103},
  [\href{https://arxiv.org/abs/1903.01473}{{\ttfamily 1903.01473}}].

\bibitem{Simpson:2022suz}
N.~Simpson and L.~Heinrich, \emph{{neos: End-to-End-Optimised Summary
  Statistics for High Energy Physics}},  3, 2022,
  \href{https://arxiv.org/abs/2203.05570}{{\ttfamily 2203.05570}},
  \href{https://doi.org/10.48550/arXiv.2203.05570}{DOI}.

\end{thebibliography}\endgroup

\end{document}